\documentclass[11pt,a4paper]{article}
\pdfoutput=1
\usepackage{jheppub}
\usepackage[utf8]{inputenc}
\usepackage{soul,xcolor}
\usepackage{cancel}
\usepackage{makecell}
\usepackage{diagbox}
\newsavebox{\foobox}

\usepackage{titlesec}
        \titleformat{\subsubsection}
           {\bfseries\normalsize}{\thesubsubsection}{1em}{}
\usepackage{ulem}
\usepackage{graphicx}
\usepackage{comment}
\usepackage{bm,array}
\usepackage{graphics}
\usepackage{epsf}
\usepackage{color}
\usepackage{amsmath}
\usepackage{amssymb}
\usepackage{latexsym}
\usepackage{slashed}
\usepackage{float}
\usepackage{cases}
\usepackage{multirow}
\usepackage{framed}
\def\ba{\begin{array}}
\def\ea{\end{array}}
\def\la{\lambda}
\def\ka{\kappa}
\def\alambda{A_\lambda}
\def\akappa{A_\kappa}
\def\mueff{\mu_\mathrm{eff}}
\def\tanb{\tan\beta}
\def\tansqb{\tan^2\beta}
\def\sQ3{\widetilde{Q}_3}
\def\sU3{\widetilde{U}_3}
\def\sD3{\widetilde{D}_3}
\def\higgsdn{H_d^0}

\def\higgsun{H_u^0}

\def\higgsdm{H_d^-}

\def\higgsup{H_u^+}

\def\mone{M_1}
\def\mtwo{M_2}
\def\mthree{M_3}

\def\mhiggsu{m_{H_u}}
\def\mhiggsd{m_{H_d}}
\def\mhiggsdsq{m_{_{H_d}}^2}
\def\mhiggsusq{m_{_{H_u}}^2}

\def\msQthree{m_{\widetilde{Q}_3}}
\def\msUthree{m_{\widetilde{U}_3}}
\def\msDthree{m_{\widetilde{D}_3}}
\def\vev{{\it vev}}
\def\vevs{{\it vevs}}
\def\vu{v_u}
\def\vd{v_d}
\def\vuplus{v_{u^+}}
\def\vdminus{v_{d^-}}
\def\vuplussq{v^2_{u^+}}
\def\vdminussq{v^2_{d^-}}
\def\vs{v_{_S}}

\newcommand{\nn}{\nonumber}
\def\veva{{\tt Vevacious}}
\newcommand{\beq}{\begin{equation}}
\newcommand{\eeq}{\end{equation}}
\newcommand{\bea}{\begin{eqnarray}}
\newcommand{\eea}{\end{eqnarray}}
\title{Spontaneous breakdown of charge in the MSSM and in the NMSSM:
       Possibilities and Implications}
\author[a, b]{Jyotiranjan Beuria}
\author[a]{AseshKrishna Datta}
\affiliation[a]{Harish-Chandra Research Institute, HBNI, Allahabad 211019, India}
\affiliation[b]{Regional Centre for Accelerator-based Particle Physics \\
                Harish-Chandra Research Institute, Allahabad 211019, India} 
\emailAdd{jyotiranjan@hri.res.in, asesh@hri.res.in}
\preprint{HRI-P-17-05-002 \\ 
\vspace*{-0.8cm}
\begin{flushright}
RECAPP-HRI-2017-006
\end{flushright}
}
\abstract{
We study the possibilities and the implications of a spontaneous breakdown of
charge in the MSSM and in the $Z_3$-symmetric NMSSM. The breakdown is triggered
by the charged states of the Higgs doublets acquiring vacuum expectation values.
In the MSSM, it is known that the presence of a charge conserving minimum for
the tree-level Higgs potential precludes a deeper (global) charge-breaking
minimum. We find that the inclusion of radiative correction to the potential
does not alter the situation while a deeper charge-conserving minimum could
arise, albeit with no major practical consequences.
In the NMSSM scenario, a charge-breaking global minimum, with or without an
accompanying charge-conserving deeper minimum, could appear even with the
tree-level Higgs potential thanks to the presence of a charge-neutral scalar
state which transforms as a singlet under the Standard Model gauge group.
Use of the NMSSM Higgs potential that includes both quantum and thermal
corrections and the requirement of a viable (stable or long-lived) vacuum that
breaks the electroweak symmetry, along with its compatibility with the latest
Higgs data, lead to nontrivial constraints on the NMSSM parameter space.
}
\begin{document} 
\setstcolor{red}
\maketitle
%
\section{Introduction}
In scenarios with two Higgs doublets, a spontaneous breakdown of charge could
occur when the charged components of the doublets acquire vacuum expectation
values (\vev). The desired (electroweak) symmetry breaking (DSB) vacuum 
conserves charge. In the presence of a spontaneous breakdown of charge, the DSB
vacuum can be, in general, accompanied by both charge-conserving (CC) and
charge-breaking (CB) minima \cite{Ferreira:2004yd}. Under such a circumstance,
a viable DSB vacuum is required to be either the global minimum of the Higgs
potential or, in case it is not (a metastable DSB vacuum), it has to have a slow
enough tunneling to the deeper CC or the CB minimum (the panic vacuum) thus
becoming cosmologically long-lived.

Crucially enough, for two Higgs doublet models (2HDM), it has been shown 
rigorously that if the tree-level Higgs potential is attributed with a CC
minimum, it has to be the deepest (global) minimum
\cite{Ferreira:2004yd, Barroso:2005sm}. 
In other words, if the potential has got a CB minimum, it can only be shallower
than the CC minimum. Such a CB minimum is also found to be invariably a saddle
point \cite{Barroso:2005sm}. Hence, if the CC minimum now happens to be the DSB
vacuum, this would be absolutely stable against tunneling to the CB minimum.
However, in the absence of a CB minimum, if there is another CC minimum apart
from the DSB one, it is to be seen if the latter still remains to be the deepest
minimum. This is since there is no general argument to prove or refute such a
possibility \cite{Barroso:2005sm}. Detailed studies of (meta)stability of the
DSB vacuum in generic 2HDM (including multi-HDM) had been taken up earlier in
references \cite{Maniatis:2006fs, Barroso:2013awa} (\cite{Barroso:2006pa}) and,
more recently, in reference \cite{Staub:2017ktc} in the context of 2HDM. In any
case, if the DSB vacuum ceases to be the global minimum, one needs to check if
it is long-lived enough so as to become viable.

Presence of additional scalars in scenarios beyond the Standard Model (SM)
invariably gives rise to more complicated scalar potentials. In some such
scenarios, one could thus naturally expect the occurrence of potential-minima
deeper than the DSB vacuum at the tree-level itself. These may destabilize the
latter as it may undergo quantum tunneling to a deeper vacuum. Requiring a
stable DSB vacuum, thus, puts stringent theoretical restrictions on the
parameter space of the scenario.
In this context, appearance of spontaneous charge and color breaking (CCB)
minima in various supersymmetric (SUSY) scenarios (as the scalar partners of the
SM quarks (squarks) and the leptons (sleptons) acquire \vev) and its
implications for the stability of the DSB vacuum have been a much-studied area
\cite{AlvarezGaume:1983gj, Gunion:1987qv, Casas:1995pd, Strumia:1996pr,
Baer:1996jn, Abel:1998cc, Abel:1998wr, Chowdhury:2013dka, Blinov:2013fta,
Hollik:2015pra, Hollik:2016dcm}. 
These general studies are only recently been followed up and improved 
\cite{Camargo-Molina:2014pwa, Chattopadhyay:2014gfa} 
within the framework of the Minimal Supersymmetric Standard Model (MSSM) by
precise treatments of several indispensable issues thus yielding a more 
conclusive picture.

Interestingly enough, inclusion of even a singlet scalar excitation (in an
otherwise 2HDM scenario) could turn the scalar potential rather nontrivial.
Early studies
\cite{Ellwanger:1996gw, Ellwanger:1999bv, Kanehata:2011ei, Kobayashi:2012xv}
in the framework of the Next-to-Minimal Supersymmetric Standard Model (NMSSM)
(which is endowed with an additional scalar which is electrically neutral and
which transforms as a singlet under the SM gauge group), though restricted in 
their scopes, uncovered some of the salient features of such a potential in
reference to the stability of the DSB vacuum. A recent in-depth study
\cite{Beuria:2016cdk} has not only lent phenomenological credence to some of
those earlier observations but has also extended the ambit of such studies by
revealing interesting, new aspects and by detailed profiling of the vacua that
appear. 

In contrast, the possibility and the implications of a spontaneous breakdown of
charge\footnote{In this work, by a `spontaneous breakdown of charge', we would
refer to such an effect triggered only by the charged Higgs states acquiring
\vevs. Charge-breaking associated with a spontaneous breakdown of
color/lepton-number and charge (CCB), as a result of the squark(s) and/or the
slepton(s) acquiring \vevs, is not considered, unless otherwise indicated.}
had attracted less attention. It may, however, be noted that such a possibility 
had earlier been pointed out \cite{Ellwanger:2009dp} in the context of the
NMSSM. Subsequently, it has been studied how such a CB minimum could appear in
the so-called Next-to-Minimal 2HDM (N2HDM) \cite{wittbrodt, Muhlleitner:2016mzt}
in which the standard 2HDM is augmented with a real singlet scalar field. The
mixing among the doublet (Higgs) and the singlet scalars induced by a
non-vanishing \vev~for the latter could then result in a CB minimum deeper than
the DSB vacuum.

Curiously enough, the possibility of a spontaneous breakdown of charge in the
minimal SUSY Standard Model (MSSM) had received even lesser attention, let alone
a thorough study of the same. To the best of our knowledge, the only (passing) 
mention of such a possibility in the MSSM context can be found in reference
\cite{Ferreira:2004yd}. The reason behind this may be the fact that, similar to
the case of a non-SUSY 2HDM scenario, a CC minimum, when it exists for the MSSM 
potential, is its global minimum, albeit at the tree-level only and when the
scalar fields in the scenario, other than the neutral and the charged Higgs
states, do not develop any \vev. Later, it was demonstrated in reference
\cite{Maniatis:2006fs} that, at the tree level, the MSSM Higgs potential could
only have the DSB vacuum as the global minimum with no accompanying local 
minimum. This happens to be a much stronger observation when compared to what
could happen in the standard 2HDM scenario discussed earlier. Given that the
MSSM is a much-constrained scenario and, in addition, the hypercharge 
assignments of the two Higgs doublets are different from that of the standard
2HDM, such an observation might not be entirely unexpected.

However, the one-loop contribution to the tree-level Higgs potential of the MSSM
could, in general, be significant because of the larger particle content of the
scenario. Thus, reference \cite{Ferreira:2004yd} pointed out that such a CC
minimum (later found to be the only minimum and which is also the DSB vacuum
\cite{Maniatis:2006fs}) could cease to remain to be the global minimum of the
radiatively-corrected potential. Instead, in principle, a CC or a CB minimum
could emerge as its global minimum. This might render the DSB vacuum unstable
with crucial implications for the regions of the MSSM parameter space that would
still remain viable. Existing literature, however, does not carry any prediction
on nature of this type of a minimum arising from such a piece of effective
Coleman-Weinberg potential \cite{Coleman:1973jx}. This is one particular area
where the present work attempts to shed light on. Furthermore, it is noted in
reference \cite{Bobrowski:2014dla} that a CC minimum deeper than the DSB vacuum
could indeed appear for a decoupled gluino and for a somewhat large value of
the higgsino mass parameter `$\mu$' when radiative corrections to the potential
arising only from the quarks and the squarks of the third generation are
considered.

The study of a spontaneous breakdown of charge in the MSSM involves at least
three scalar fields (two neutral and one charged components of the doublet Higgs
fields) developing \vevs. Note that a suitable set of \vevs~for the neutral
(doublet) Higgs fields is always required to ensure the desired breaking of the
electroweak symmetry. In the $Z_3$-symmetric NMSSM, in addition, one needs a
nonvanishing \vev~for the singlet scalar field ($S$) as well that dynamically
gives rise to the `$\mu$' parameter, $\mueff$, thus solving the well-known
``$\mu$-problem'' \cite{Kim:1983dt}. 

In the presence of the singlet scalar field `$S$', a CB minimum could turn out
to be the global minimum of the $Z_3$-symmetric NMSSM potential, already at the
tree-level. This is in sharp contrast to the MSSM case discussed earlier.
However, finding all the minima and hence determining the global one (which is
crucial for the purpose) in a situation where multiple scalar states could
acquire \vevs~is expected to be a non-trivial exercise. The problem has earlier
been approached analytically in reference \cite{Maniatis:2006jd}. The task
becomes even harder when radiative corrections are to be necessarily included.
To complicate things further, the \vevs~for the charged Higgs fields (that
trigger a breakdown of charge-conservation) induce mixing among the
fermions/sfermions \cite{Eberl:1993xu}, the notables ones being between the top
and the bottom quarks and among the top and the bottom squarks. Such mixings, in
turn, affect the radiative corrections to the Higgs potential. Furthermore, in
the presence of a deeper CB minimum, one needs to check the stability of the DSB
vacuum against its tunneling to the former.

For an optimal handling of such a set of rather involved tasks, one needs to
resort to a numerical approach to the problem. The package \veva~{\tt (v1.2.02)}
\cite{Camargo-Molina:2013qva} provides us with such an elaborate computing
framework. \veva~uses the principle of homotopy continuation via the package
{\tt HOM4PS2} \cite{hom4ps2} for an exhaustive hunt for all possible minima of
the supplied potential. It further incorporates full 1-loop corrected effective
potential using inputs from {\tt SARAH (v4.12.1)}
\cite{Staub:2013tta, Staub:2015kfa}-generated {\tt SPheno}
\cite{Porod:2003um, Porod:2011nf} package. The package
{\tt CosmoTransitions (v2.0.02)} \cite{Wainwright:2011kj} is employed from
within \veva~to estimate the tunneling time of the DSB vacuum to a possible
deeper minimum.

Recently, some salient aspects and implications of a spontaneous breakdown of
charge in the NMSSM scenario have been discussed in the literature
\cite{Krauss:2017nlh}\footnote{This work came out while we had been halfway
through the present study.} using \veva. The present study performs a thorough
scan of the relevant parameter space using \veva. It benefits from and broadly
agrees with some specific observations made in reference \cite{Krauss:2017nlh},
within the scopes mentioned there, and extends beyond to obtain a detailed
understanding of the phenomenon in the $Z_3$-symmetric NMSSM. Furthermore, we
also undertake a detailed study of the MSSM scenarios with a similar goal.

\veva~also has the provision to consider the finite temperature (thermal)
effects to the potential which, in general, cannot be ignored
\cite{Affleck:1980ac, Linde:1980tt, Linde:1981zj, Brignole:1993wv}. Some recent
studies have concretely established its important role in deciding the fate of
the DSB vacuum \cite{Camargo-Molina:2014pwa, Beuria:2016cdk}. We include the
thermal contribution to the potential in our present study, at length. We also
subject our scans to the latest experimental constraints from the observed Higgs
sector by using packages like {\tt HiggsSignals (v1.4.0)} \cite{Bechtle:2013xfa}
and {\tt HiggsBounds (v4.3.1)} \cite{Bechtle:2013wla}. In particular, a scenario
like the NMSSM, which allows for mixing among the neutral doublet Higgs states
and the singlet scalar, is naturally much sensitive to these constraints.

The paper is organized as follows. In section \ref{sec:mssm-cb} we first take up
an analytical study of various flat directions in the MSSM field space to check
if a CB (and/or a CC) minimum deeper than the DSB vacuum could appear for the
tree-level Higgs potential. This is followed by a scan of the MSSM parameter
space using \veva~which incorporates both quantum and thermal corrections to the
Higgs potential. We also present a corroborative study based entirely on an
alternate spectrum generator like {\tt FeynHiggs (v2.13.0)}
\cite{Heinemeyer:1998yj, Heinemeyer:1998np, Degrassi:2002fi, Frank:2006yh,
Hahn:2013ria, Bahl:2016brp} 
and our dedicated {\tt Mathematica} \cite{mathematica} routine that is used for
the analysis. Thus, we delineate the regions of the MSSM parameter space where a
minimum deeper than the DSB vacuum appears and indicate its implications for the
stability of the latter. Section \ref{sec:nmssm-cb} presents an analytical study
of various flat directions in a more involved field space of the NMSSM along
which a deeper CC and/or a CB minimum could appear. This is again followed by a
dedicated search for such deeper minima using \veva~and then finding if these
are of the CC or CB types and further reflecting on how critical they could be
to the stability of the DSB vacuum. The role of thermal correction to the
potential is discussed. All through, regions compatible with an SM-like Higgs
boson are indicated. In section \ref{sec:conclusions} we conclude.
%
\section{Spontaneous breakdown of charge: the MSSM case}
\label{sec:mssm-cb}
%
As pointed out in the Introduction, the tree-level Higgs potential of the MSSM,
a SUSY variant of a generic 2HDM scenario, has an in-built robust protection 
against developing a CB minimum deeper than the DSB (CC) vacuum, when the latter
is present \cite{Ferreira:2004yd, Barroso:2005sm}. However, it remains to be
seen if a deeper minimum could arise (and its nature (CB or CC or both)) when
radiative correction to the potential is included. Natural directions along
which this might happen are the so-called flat directions for which the
tree-level potential already possesses
minima\footnote{Directions along which the tree-level
potential is unbounded from below could also develop a minimum when radiative
correction is included \cite{Bobrowski:2014dla}.}.
In the following, we first study such flat directions of the tree-level Higgs
potential analytically and explore if these could give rise to deeper CB/CC
minima. This is followed by a general study of such a phenomenon via numerical
means using the {\tt SPheno}-\veva~and the {\tt FeynHiggs-Mathematica}
frameworks discussed in the Introduction. Both the frameworks incorporate the
full 1-loop correction to the Higgs (scalar) potential.
%
\subsection{Analysis of the tree-level Higgs potential: the MSSM case}
\label{subsec:analysis-mssm}
%
The Higgs potential involving both neutral and charged Higgs fields is given by
%
\bea
V_\mathrm{Higgs} & = & \left(m_{H_u}^2 + \left| \mu\right|^2\right) 
\left(\left|H_u^0\right|^2 + \left|H_u^+\right|^2\right) 
+\left(m_{H_d}^2 + \left| \mu \right|^2\right) 
\left(\left|H_d^0\right|^2 + \left|H_d^-\right|^2\right) \nn \\
&&+\frac{g_1^2+g_2^2}{8}\left(\left|H_u^0\right|^2 +
\left|H_u^+\right|^2 - \left|H_d^0\right|^2 -
\left|H_d^-\right|^2\right)^2
+\frac{g_2^2}{2}\left|H_u^+ H_d^{0*} + H_u^0 H_d^{-*}\right|^2\nn \\
&&+B_{\mu} (H_u^+ H_d^- -H_u^0 H_d^0)+ \mathrm{h.c.} \; ,
\label{eq:Vmssm}
\eea
%
where $\mhiggsusq$ and $\mhiggsdsq$ are the soft masses for the $u$- and the
$d$-type Higgs excitations, $g_1$ and $g_2$ are the $U(1)_Y$ and $SU(2)_L$ gauge
couplings and $B_\mu$ is the soft term corresponding to the $\mu$-term in the
MSSM superpotential. Note that successful electroweak symmetry breaking (EWSB)
requires $B_\mu >0$ given our convention of $\tanb >0$, where
$\tanb={\vu \over \vd} (> 1)$, the ratio of the \vevs~of the neutral components
of the two Higgs doublets. The tadpole conditions corresponding to these Higgs
fields are given by
%
\begin{subequations}
\label{eq:tadpoles-mssm}
\bea
T_{\higgsun} =
\frac{\partial V_{\mathrm{Higgs}}}{\partial v_u} = 0 &=& g^2
v_u(v_u^2+v_{u^+}^2-v_d^2) + 2 (m_{H_u}^2+\mu ^2)v_u -2 B_\mu v_d \nn \\
&&+ \, v_{d^-}(\frac{g_2^2-g_1^2}{2} v_u v_{d^-}+ g_2^2 v_d v_{u^+}) \: ,
\label{eq:Tu}
\\ [8pt]
T_{\higgsdn} =
\frac{\partial V_{\mathrm{Higgs}}}{\partial v_d} =0 &=& g^2
v_d(v_d^2+v_{d^-}^2-v_u^2) + 2 (m_{H_d}^2+\mu ^2)v_d -2 B_\mu v_u  \nn \\
&&+ \, v_{u^+}(\frac{g_2^2-g_1^2}{2} v_d v_{u^+}+ g_2^2 v_u v_{d^-}) \: ,
\label{eq:Td}
\\ [8pt]
T_{\higgsup} =
\frac{\partial V_{\mathrm{Higgs}}}{\partial v_{u^+}} =0 &=& g^2
v_{u^+}(v_u^2+v_{u^+}^2-v_{d^-}^2) + 2 (m_{H_u}^2+\mu ^2)v_{u^+} +2 B_\mu
v_{d^-}  \nn \\
&&+ \, v_d(\frac{g_2^2-g_1^2}{2} v_d v_{u^+}+ g_2^2 v_u v_{d^-}) \: ,
\label{eq:Tup}
\\ [8pt]
T_{\higgsdm} =
\frac{\partial V_{\mathrm{Higgs}}}{\partial v_{d^-}} =0 &=& g^2
v_{d^-}(v_d^2+v_{d^-}^2-v_{u^+}^2) + 2 (m_{H_d}^2+\mu ^2)v_{d^-} +2 B_\mu
v_{u^+}  \nn \\
&&+ \, v_u(\frac{g_2^2-g_1^2}{2} v_u v_{d^-}+ g_2^2 v_d v_{u^+}) \: ,
\label{eq:Tdm}
\eea
\end{subequations}
%
where $\vu$, $\vd$, $\vuplus$ and $\vdminus$ are all considered to be real and
represent constant field values of the respective fields only at a minimum of
the potential (i.e., the \vevs). Equations \ref{eq:Tu} and \ref{eq:Td} could be
solved for $\mhiggsusq$ and $\mhiggsdsq$ at the DSB vacuum, i.e., when
$\vu=v_0\sin\beta$ and $\vd=v_0\cos\beta$, where $v_0$ is the overall Higgs 
\vev~($v_0=174$ GeV) with which the electroweak symmetry is broken. Thus, one
finds
%
\begin{subequations}
\label{eq:soft-mass2}
\bea
m_{H_u}^2 &=& B_\mu \cot \beta +\frac{1}{4} {v_0}^2 \cos (2 \beta )
\left({g_1}^2+{g_2}^2\right)-\mu ^2 \; ,  \\ 
\label{eq:soft-mass2-hu}
m_{H_d}^2&=& B_\mu \tan \beta -\frac{1}{4} {v_0}^2 \cos (2 \beta )
\left({g_1}^2+{g_2}^2\right)-\mu ^2 \; .
\label{eq:soft-mass2-hd}
\eea
\end{subequations}
%
The DSB vacuum preserves charge. Hence $\vuplus=\vdminus=0$ at the DSB vacuum.
Also, unless otherwise specified, throughout this work, $\vu$, $\vd$, $\vuplus$
and $\vdminus$ would stand for generic \vevs~for the respective Higgs fields.
The depth of the Higgs potential at the DSB vacuum can now be found by
substituting $\mhiggsusq$ and $\mhiggsdsq$ from equation \ref{eq:soft-mass2}
into equation \ref{eq:Vmssm} and is given by
%
\bea
\label{eq:depth-dsb}
V_\mathrm{Higgs}^{\mathrm{DSB}} &=& \frac{-g^2}{4} v_0^4 \cos^2 2\beta \; ,
\eea
%
where $g^2={{g_1^2 + g_2^2} \over 2}$. A similar substitution but allowing also
for nonvanishing \vevs~for the charged Higgs fields yields the depth of the
Higgs potential at a possible non-DSB (\cancel{DSB}) minimum and is given by
%
\bea
V_\mathrm{Higgs}^\mathrm{\cancel{DSB}} & = & (B_\mu \cot\beta +\frac{g^2}{2} v_0^2 \cos2\beta) (v_u^2+v_{u^+}^2)+  (B_\mu \tan\beta -\frac{g^2}{2} v_0^2 \cos2\beta) (v_d^2+v_{d^-}^2)  \nn \\
&+& \frac{g^2}{4} (v_u^2+v_{u^+}^2-v_d^2-v_{d^-}^2)^2  +2B_\mu (v_{u^+}v_{d^-}-v_u v_d) + \frac{g_2^2}{2} (v_{u^+}v_d  + v_u v_{d^-})^2 \:\: .
\label{eq:depth-cb}
\eea
%
At this point, one needs to exercise caution before associating the
non-vanishing \vevs~for the charged Higgs states to a breakdown of charge. 
Since we are working in a 4-\vev~framework, there is always an $SU(2)$ rotation which
one could apply simultaneously to both Higgs doublets. 
Note that this issue is generic to a 2HDM setup without 
a direct reference to the potential derived in equation \ref{eq:depth-cb}\footnote{
This is also true for the NMSSM case studied later in section \ref{sec:nmssm-cb}.}. For
the particular hypercharge asisgnments for the Higgs doublets as in SUSY 2HDM, the rotated (by an angle $\theta$) configurations for the set of \vevs~
$
\left(
\begin{array}{l}
\vuplus \\ \vu
\end{array}
\right)
$
and
$
\left(
\begin{array}{l}
\vd \\ \vdminus
\end{array}
\right)
$
can then be given by
\bea
\left(
\begin{array}{l}
\vuplus^\prime \\ \vu^\prime
\end{array}
\right) =
\left(
\begin{array}{l}
\vuplus \cos \theta - v_u \sin \theta \\
\vuplus \sin \theta + v_u \cos \theta
\end{array}
\right)
\quad \mathrm{and} \quad
\left(
\begin{array}{l}
\vd^\prime \\ \vdminus^\prime
\end{array}
\right) =
\left(
\begin{array}{l}
v_d \cos\theta - \vdminus \sin\theta \\
v_d \sin \theta + \vdminus \cos \theta
\end{array}
\right) \; .
\label{eq:field-rotation}
\eea
%
Thus, one could always find a value of `$\theta$' which rotates away one of the
charged \vevs~in the new basis \cite{Barroso:2006pa}\footnote{In this
work we would continue to consider \vevs~for both the charged states
explicitly.}. By choosing $\vuplus^\prime=0$, we find
%
\bea
v_u^\prime= \sqrt{\vu^2+\vuplussq}, \quad 
v_d^\prime=\frac{\vu \vd - \vuplus \vdminus}{\sqrt{\vu^2+\vuplussq}}
\quad \mathrm{and} \quad
\vdminus^\prime= \frac{\vuplus \vd + \vu \vdminus}{\sqrt{\vu^2+\vuplussq}} \; .
\label{eq:field-rotation2}
\eea
%
If vacua deeper than the DSB one were to appear, these will be most likely along
some ($D$-) flat directions in the field space which, from equation
\ref{eq:depth-cb}, are given by
%
\begin{subequations}
\label{eq:flat-dir}
\beq
\vuplus \vd +  \vu \vdminus = 0 \; ,
\label{eq:flat-dir-mssm-1}
\eeq
\beq
\vu^2 + \vuplussq - \vd^2 - \vdminussq = 0 \; . 
\label{eq:flat-dir-mssm-2}
\eeq
\end{subequations}
%
Note that along the $D$-flat direction of \ref{eq:flat-dir-mssm-1},
$\vdminus^\prime$ in equation \ref{eq:field-rotation2} vanishes. Thus, along
such a direction both charged \vevs~get simultaneously rotated away. Hence the
minimum of the potential does not break charge even if, in the original basis,
non-vanishing \vevs~appear explicitly in the potential. This should be
corroborated by a vanishing photon mass that such a configuration of \vevs~leads
to and which serves as a robust pointer to the phenomenon of charge conservation.
In fact, any CC minimum (including the DSB minimum) has to exist only along this
flat direction of equation \ref{eq:flat-dir-mssm-1} when charged \vevs~are
incorporated. However, it still remains to be seen if such a direction could
give rise to a deeper CC minimum. 

To this end, the $D$-flat direction of equation \ref{eq:flat-dir-mssm-2} may be
explored in conjunction. Requiring a CC minimum along this direction would
relate the \vevs~further. These relations can be found by imposing the necessary
condition of \ref{eq:flat-dir-mssm-1} for having a CC minimum on the \vevs~
appearing in \ref{eq:flat-dir-mssm-2} and are given by the following set of
conditions:
%
\beq
v_u= \pm v_d, \;\; \vuplus=\mp \vdminus \; . 
\eeq
%
%
%
%
In the MSSM context, in particular, this implies the trivial solution
$v_u= v_d=\vuplus=\vdminus =0$ which corresponds to a minimum shallower than 
the DSB vacuum.
Can a CB minimum appear along the $D$-flat direction of equation
\ref{eq:flat-dir-mssm-2}? A priori, this cannot be ruled out. Given that the
minimal number of \vevs~required for a CB minimum to exist is two (one neutral
\vev, along with a charged \vev), we may expect to find such a CB minimum by
choosing $\vuplus^\prime=v_d^\prime=0$ or $\vdminus^\prime = v_u^\prime=0$.
The first choice can be directly plugged into the expression of $v_d^\prime$
in equation \ref{eq:field-rotation2}. Similarly, the second choice would work
with an expression for $v_u^\prime$ from a set analogous to equation
\ref{eq:field-rotation2} that can be found by choosing $\vdminus^\prime=0$
instead. In either case, this would result in
%
\beq
v_u v_d = \vuplus \vdminus \; .
\label{eq:cb-flat}
\eeq 
%
Equation \ref{eq:cb-flat} in conjunction with \ref{eq:flat-dir-mssm-2} leads to
%
\beq
\vuplus= \pm v_d ,\quad \vdminus=\pm v_u \; .
\eeq
%
As for the MSSM case, these relations are 
not compatible with the corresponding tadpole conditions.
Thus, at the tree-level, the MSSM Higgs potential cannot have a CB extremum.
This is in agreement with the findings in reference \cite{Maniatis:2006fs}.

Again in the MSSM context,
we turn back to see if the $D$-flat direction given by equation
\ref{eq:flat-dir-mssm-1}, which can always give rise to a CC minimum, could, by
itself, develop a deeper one this time. 
Using equations \ref{eq:soft-mass2} and \ref{eq:flat-dir-mssm-1} one could
simplify the set of tadpoles given in equations \ref{eq:tadpoles-mssm}. 
In a rotated $vev$ configuration with neutral $vevs$ ($v_{u^+}^\prime=v_{d^-}^\prime=0$),
the tadpoles for $\higgsun$ (equation \ref{eq:Tu}) and $\higgsdn$ (equation
\ref{eq:Td}) then reduce, respectively, to
%
\begin{subequations}
\label{eq:reduced-conditions-1-2}
%
\beq
\label{eq:condition-1}
{\vu^\prime}^2 - {\vd^\prime}^2  +v_0^2 \cos2\beta ={2 \over g^2}
B_\mu(\frac{v_d^\prime}{v_u^\prime}-\cot\beta) \: ,
\eeq 
%
\beq
\label{eq:condition-2}
{\vu^\prime}^2 - {\vd^\prime}^2  +v_0^2 \cos2\beta ={2 \over g^2}
B_\mu(\tan\beta - \frac{v_u^\prime}{v_d^\prime}) \;. 
\eeq
\end{subequations}
%
Given the identical expressions on the left hand side of these equations, two 
solutions for ${\vu \over \vd}$ ($=\tan\beta, \, \cot\beta$) turn out to be 
consistent. However, the solution ${\vu^\prime \over \vd^\prime} = \cot\beta$ gives rise to
complex \vevs. We ignore this solution since it is in conflict with our 
original assumption. Hence using the solution ${\vu^\prime \over \vd^\prime} =\tan\beta$ 
with either of equations \ref{eq:condition-1} or \ref{eq:condition-2}, we find
%
\beq
{\vu^\prime }^2 = v_0^2 \sin^2\beta \qquad  \mathrm{and} \qquad  
{\vd^\prime}^2= v_0^2 \cos^2\beta \: ,
\label{eq:vevs-mssm}
\eeq
%
thus leading to
%
\beq
{\vu^\prime }^2 + {\vd^\prime }^2 = v_0^2 \:,
\label{eq:locus-mssm}
\eeq
%
This is exactly the DSB vacuum obtained in equation \ref{eq:depth-dsb}. 
This is again
in agreement with the findings of reference \cite{Maniatis:2006fs} which
indicates that the DSB vacuum, when present, is the global minimum of the
tree-level MSSM Higgs potential.
At this point, it is interesting to note that, had we continued to work
with all four $vevs$, we would have ended up with an infinite number of vacua with
non-vanishing charged \vevs, which are all identical to the DSB vacuum connected
via $SU(2)$ symmetry we discussed earlier. We will discuss its artifact at the 
end of next subsection in the context of a {\tt Vevacious} analysis.
%
\subsection{Scanning of the MSSM parameter space}
\label{subsec:scan-mssm}
%
In this section, we undertake a numerical study that sheds light on the regions
of the MSSM parameter space with viable DSB vacuum when only the Higgs fields
could acquire \vevs. The dedicated package \veva~is used for the purpose which,
in turn, uses the full 1-loop corrected effective potential with input
parameters taken from {\tt SARAH}-generated {\tt SPheno}. Since \veva~employs a
radiatively corrected Higgs potential, it might be able to explore subtle and
potentially crucial effects which do not show up with the tree-level potential
that we adhered to in our analytical study in section \ref{subsec:analysis-mssm}.
Thus, it would be interesting to see if a deeper minimum for the Higgs potential
(of either a CC- or a CB-type) appears having immediate implications for the
stability of the DSB vacuum.

Furthermore, it has been correctly pointed out in reference \cite{Krauss:2017nlh}
that it is not entirely justified to assign the deeper minimum closest in the
field space to the DSB vacuum to be the panic vacuum, only to which tunneling of
the former is considered, as is the case for the publicly available version of
\veva. Accordingly, we tweak \veva~to check all deeper minima to find the panic
vacuum as the most dangerous one (with the fastest tunneling time) of them all.
In addition, as a corroborative measure, we use {\tt FeynHiggs} to generate the
MSSM spectra and employ our dedicated {\tt Mathematica} routine to minimize the
full 1-loop corrected potential. The analysis is further subjected to the
constraints coming from the observed Higgs sector via the use of packages like
{\tt HiggsSignals} and {\tt HiggsBounds}.
%
%
\begin{figure}[t]
\centering
\includegraphics[height=0.25\textheight,width=0.40\columnwidth,clip]{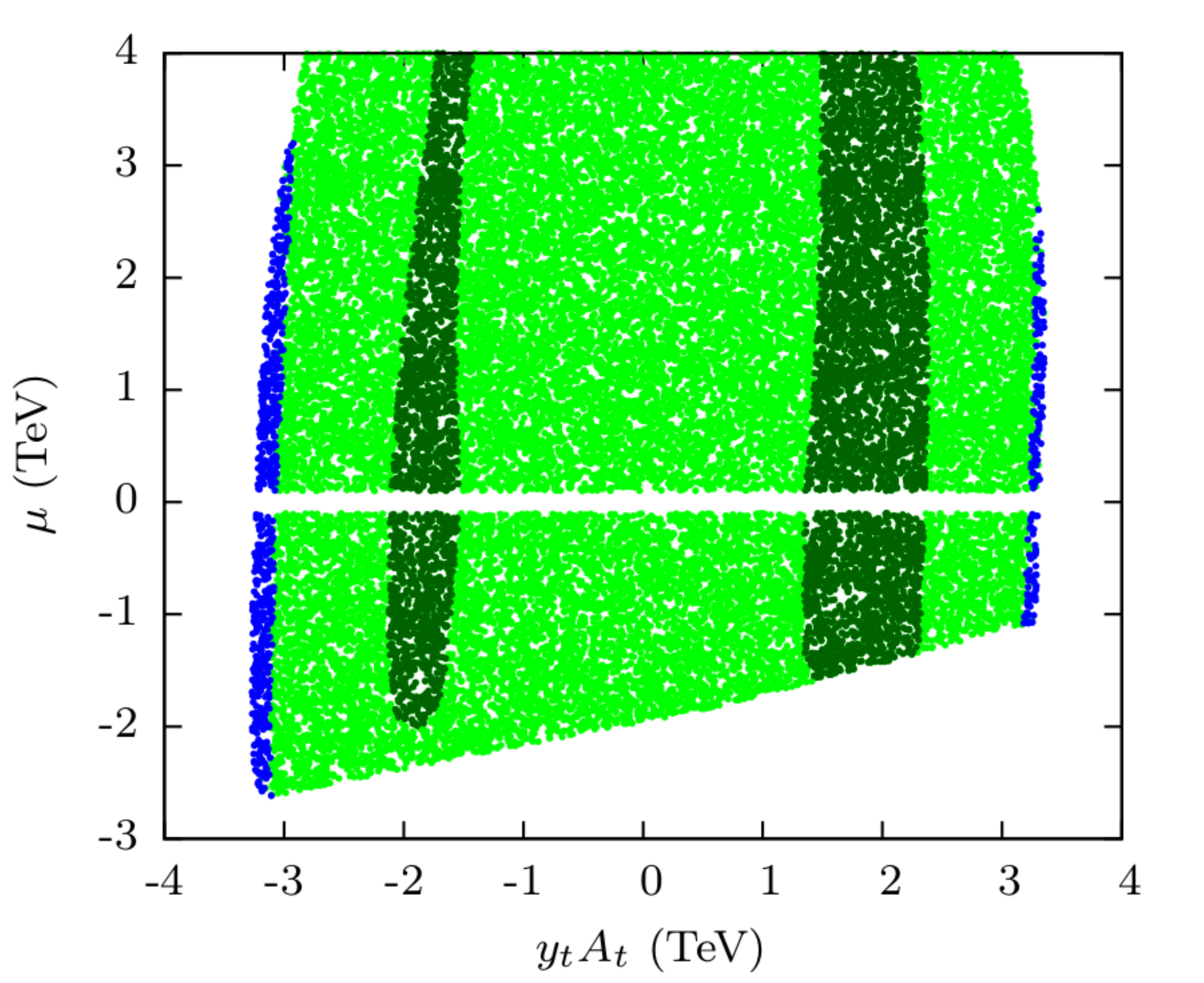}
\includegraphics[height=0.25\textheight,width=0.40\columnwidth,clip]{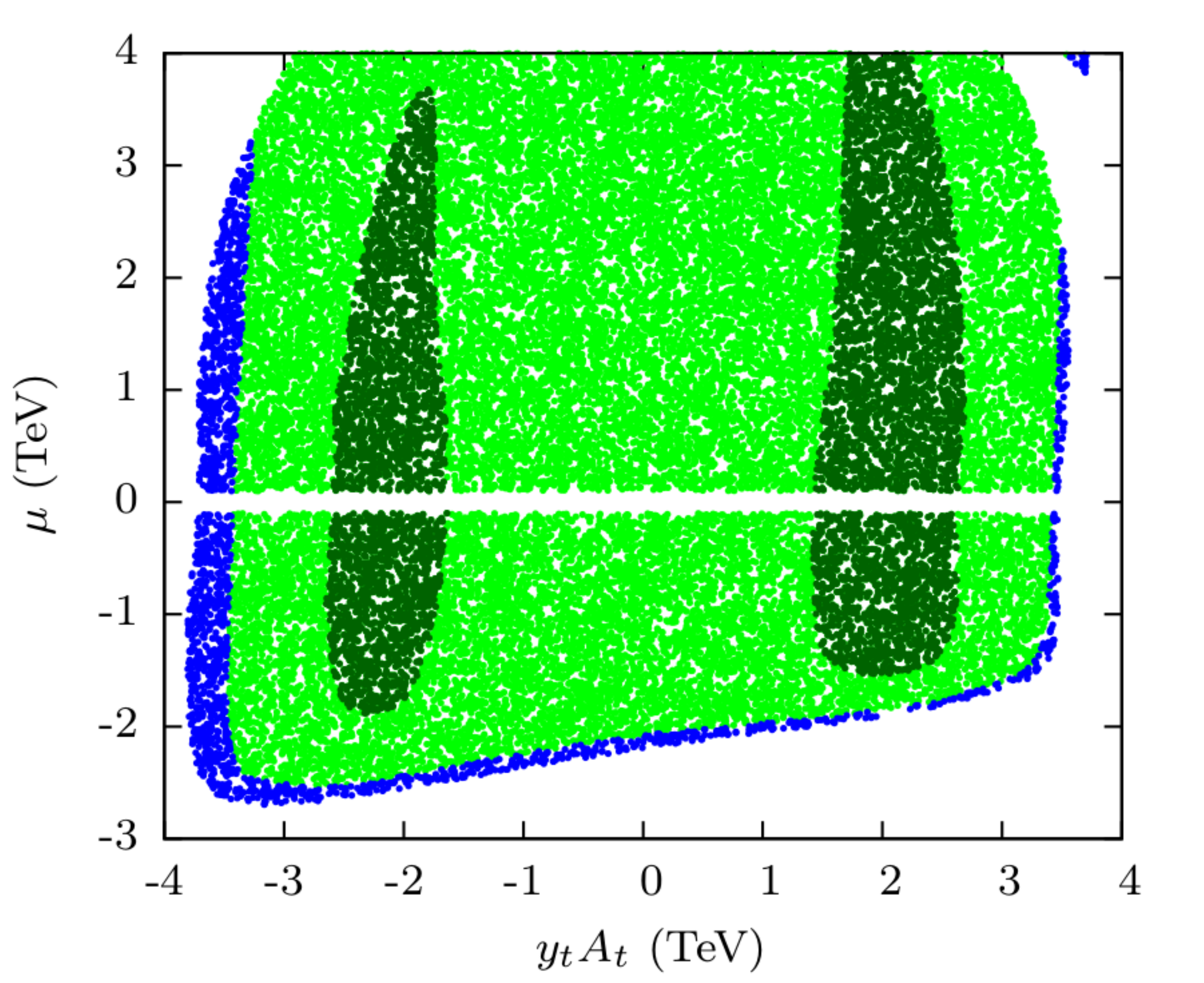}
\includegraphics[height=0.25\textheight,width=0.40\columnwidth,clip]{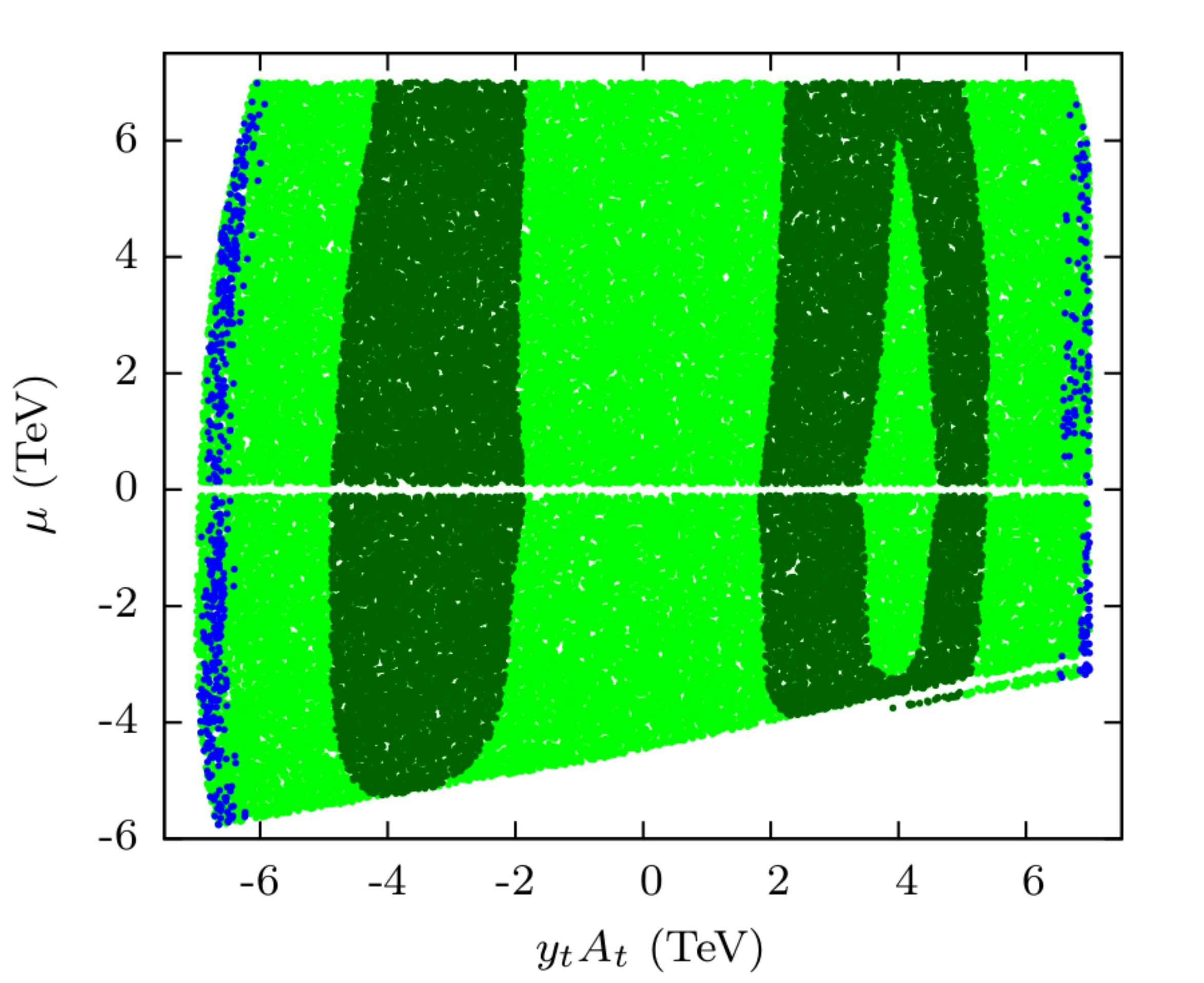}
\includegraphics[height=0.25\textheight,width=0.40\columnwidth,clip]{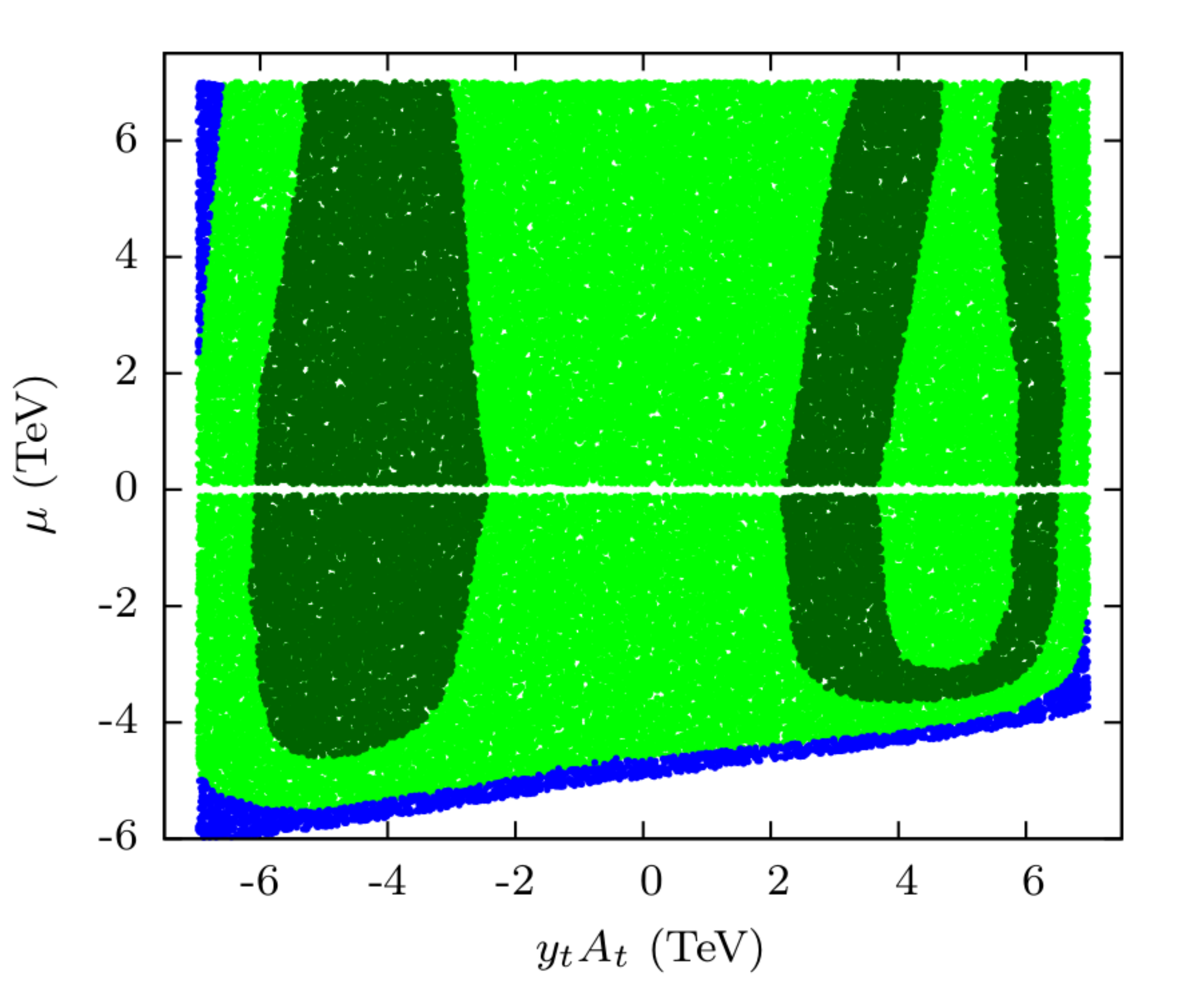}
\caption{
Scatter plots showing the regions with the DSB vacuum as the global minimum of
the potential (green) and with an accompanying non-DSB CC minimum as the global
minimum of the same (in blue) in the $y_t A_t$--$\mu$ plane. Plots in the left
(right) panel are obtained from the
{\tt SPheno}-\veva~({\tt FeynHiggs-Mathematica}) framework. The top (bottom)
panel corresponds to
$\msQthree=\msUthree=\msDthree= 1 \, \mathrm{TeV} \, (2$ TeV). Other fixed
parameters are as follows: $\mone=\mtwo=750 \, \mathrm{GeV}$ and $\mthree=2$
TeV, $m_A=2$ TeV, $A_b=0$ and $\tanb=25$. The renormalization scale is set to
$Q=\sqrt{\msQthree \msUthree}$. Regions in dark-green are compatible with the
observed SM-like Higgs boson (at $\sim 95\%$ C.L.) as reported by
{\tt HiggsBounds} and {\tt HiggsSignals}.
}
\label{fig:mssm-mu-At}
\end{figure}
%
The results of random scans over a relevant set of MSSM parameters showing the 
stability pattern of the DSB vacuum are presented in figure \ref{fig:mssm-mu-At}.
Given that the top squark sector is expected to dominate in the radiative
contributions to the potential, we choose the $y_t A_t$--$\mu$ plane for
illustration. In each row, the left plot results from a \veva~analysis of the
spectra obtained from {\tt SARAH}-generated {\tt SPheno} where we indicate the
regions that correspond to either a stable DSB vacuum (global minimum; in green)
or the presence of an accompanying (non-DSB) global minimum (in blue). The
corresponding right plots present the results of a similar analysis using the
same set of MSSM input parameters but adopting the {\tt FeynHiggs-Mathematica}
framework. The top (bottom) panel corresponds to
$\msQthree=\msUthree=\msDthree=1 \, \mathrm{TeV} \, (2\, \mathrm{TeV})$.
Ranges of various MSSM parameters that are scanned over and the fixed values for
the others are indicated in the figure caption.

No CB minimum emerges, irrespective of whether it is deeper than the DSB vacuum
or not. However, a deeper (panic) CC minimum, which is absent for the tree-level
Higgs potential, might appear this time with its origin in the radiative
correction to the said potential (Coleman-Weinberg type). Such deeper CC minima,
however, appear only along the edges (in blue) of the displayed plane. A similar
phenomenon associated with larger values of `$\mu$', along with a decoupled
gluino, has been observed in reference \cite{Bobrowski:2014dla} which
incorporates corrections to the potential from the third generation quarks and
squarks only. Interestingly, as can be gleaned from figure \ref{fig:mssm-mu-At},
the inclusion of the full 1-loop correction to the potential (as is the case
with both {\tt SPheno} and {\tt FeynHiggs}) results in such a panic CC minimum
occurring even for relatively smaller values of `$\mu$' and $A_t$. However, such
regions of the MSSM parameter space appear to be not compatible with the
observed mass of the SM-like Higgs boson (given by the dark-green bands). This
(mostly) pre-empts the threat from an emerging deeper CC minimum destabilizing
the DSB vacuum. The horizontal, blank stripes about $\mu=0$ indicate the ranges
of unacceptable $\mu$-values dictated by experimental constraints, primarily
from the chargino searches.

\begin{figure}[t]
\centering
\includegraphics[height=0.33\textheight,width=0.49\columnwidth,clip]{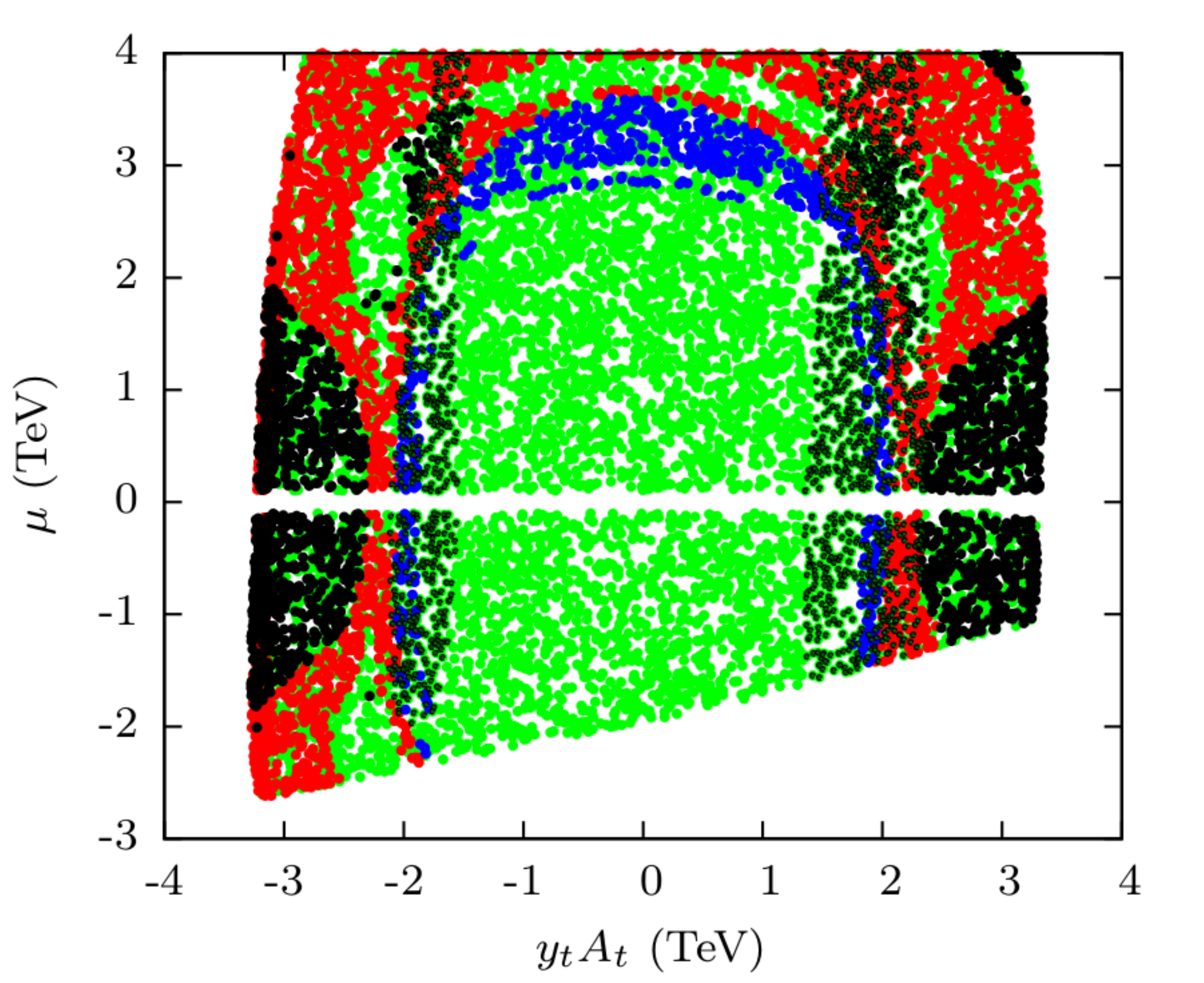}
\includegraphics[height=0.33\textheight,width=0.49\columnwidth,clip]{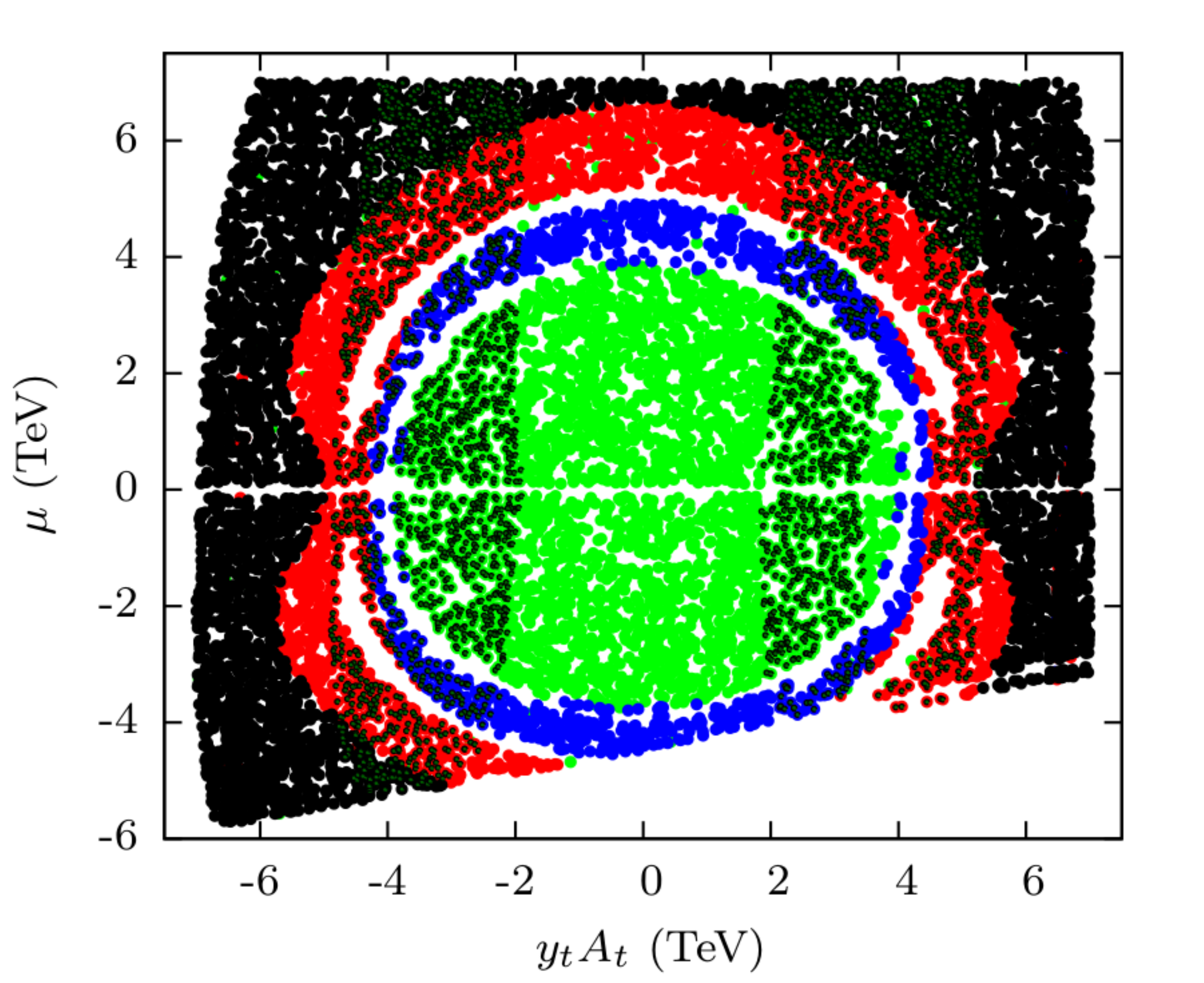}
\caption{
Stability status of the DSB vacuum in the presence of a deeper CC vacuum along
with an unavoidable deeper CCB vacuum for the cases presented in the left panel 
({\tt SPheno-\veva} analyses) of figure \ref{fig:mssm-mu-At}. The left (right)
plot corresponds to the top left (bottom left) plot of figure
\ref{fig:mssm-mu-At}. The color code adopted is as follows: green stands for a
stable DSB vacuum (global minimum), blue represents a metastable but
cosmologically long-lived DSB vacuum, black indicates the presence of a DSB
vacuum which is unstable under quantum tunneling at zero temperature while red
corresponds to a DSB vacuum unstable against tunneling when the finite
temperature corrections to the potential are included. The vertical bands in
dark-green again delineate the regions compatible with the observed SM-like
Higgs boson.
}
\label{fig:mssm-mu-At-ccb}
\end{figure}
\begin{figure}[h]
\centering
\includegraphics[height=0.40\textheight, width=0.49\columnwidth ,
clip]{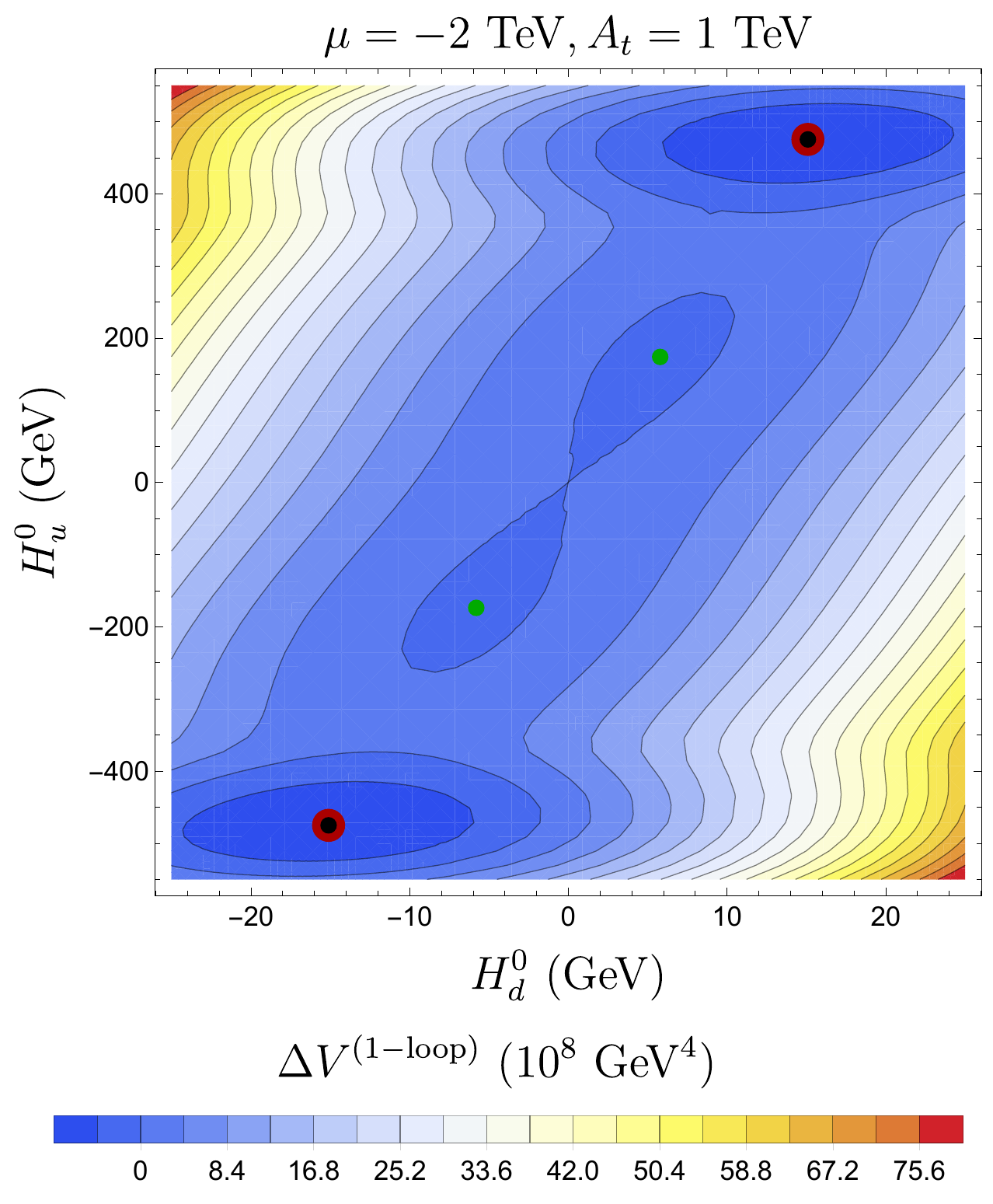}
\caption{
Potential ($\Delta V$) contours in the $H_d^0$--$H_u^0$ plane showing the
(equivalent) locations of the DSB vacuum (blobs in green) and the accompanying
global CC minimum (blobs in black-within-red). The magnitudes of $\Delta V$
can be estimated from the color-palette displayed underneath. The fixed MSSM
parameters `$\mu$' and $A_t$ are as indicated at the top of the plot. Other
fixed parameters are as in figure \ref{fig:mssm-mu-At} with
$\msQthree=\msUthree=\msDthree=1$ TeV. See text for details.
}
\label{fig:mssm-contour}
\end{figure}
%

In figure \ref{fig:mssm-mu-At-ccb}, we indicate the stability status of the DSB
vacuum for the same set of data points and in the same parameter plane as for
figure \ref{fig:mssm-mu-At} but this time we allow for the colored sfermions
assuming \vevs. This is to check for likely appearances of deeper CCB vacua that
could have already been triggered by the large values of $y_t A_t$ and/or
`$\mu$' that are necessary for a deeper CC vacuum to occur, as is seen in figure
\ref{fig:mssm-mu-At}. 

We indeed find an onset of a deeper CCB minimum just away from the central
regions of these plots with much smaller values of $y_t A_t$ and `$\mu$'. This
clearly indicates that long before a deeper CC minimum originating purely in the
Higgs potential could pose a threat to the stability of the DSB vacuum, the
scalar potential is inflicted with a dangerous CCB minimum. This may perhaps be
easily comprehended by noting that both deeper CCB and CC minima are dominantly
driven by the fermions and sfermions from the third generation; however, while
the former could well be a tree-level effect, the latter have a genuine origin
in the radiatively corrected potential.
%
%

In figure \ref{fig:mssm-contour} we demonstrate the potential ($\Delta V$)
contours in the $H_d^0$--$H_u^0$ plane, where $\Delta V$ is the relative depth 
with respect to the potential at the field origin. The location(s) of the DSB
vacuum (deeper minimum) are indicated by the green (black-in-red) blobs. Such
minima appearing in the first quadrant are the identical ones to those showing
up in the third quadrant because of the underlying reflection symmetry of the
potential.

Before concluding this section, one observation regarding occasional
numerical (in)stability of the results obtained from \veva~ may be noted.
While working with two non-vanishing charged Higgs \vevs, 
one encounters a CC minimum nearly degenerate with the DSB
vacuum and situated very close to the latter in the field space. Under such
a circumstance, due to limited floating point precision it uses, \veva~
may find it difficult to decide the fate of the DSB vacuum correctly.
However, this is a direct consequence of not exploiting the freedom to
rotate one of the charged Higgs $vevs$ to start with.
We further checked this via our
{\tt Mathematica} analysis using a much larger precision that it offers.
%
\section{Spontaneous breakdown of charge: the $Z_3$-symmetric NMSSM case}
\label{sec:nmssm-cb}
%
It has recently been noted \cite{Muhlleitner:2016mzt} in the context of a
singlet-extended 2HDM (N2HDM) that the global minimum of the potential may not
be a charge-conserving one at the tree-level, unlike in the 2HDM. This is
attributed to the neutral singlet scalar field of such a scenario developing
\vev~thereby mixing with the doublet Higgs states. Naturally, such an
observation bears relevance to a scenario like the NMSSM where a similar effect
can be investigated \cite{Maniatis:2006jd}. In this section, we first take an
analytical look into how the $Z_3$-symmetric NMSSM Higgs potential could develop
a global CB minimum. A numerical, random scan of the NMSSM parameter space using
\veva~follows. This delineates the region of the parameter space offering
a viable DSB vacuum.
%
\subsection{Analysis of the tree-level  Higgs potential: the NMSSM case}
\label{subsec:analysis-nmssm}
%
The tree-level Higgs potential of the $Z_3$-symmetric NMSSM is given by
%
\bea
V_\mathrm{Higgs} & = & \left|\lambda \left(H_u^+ H_d^- - H_u^0
H_d^0\right) + \kappa S^2 \right|^2 \nn \\
&&+\left(m_{H_u}^2 + \left| \lambda S\right|^2\right) 
\left(\left|H_u^0\right|^2 + \left|H_u^+\right|^2\right) 
+\left(m_{H_d}^2 + \left| \lambda S\right|^2\right) 
\left(\left|H_d^0\right|^2 + \left|H_d^-\right|^2\right) \nn \\
&&+\frac{g_1^2+g_2^2}{8}\left(\left|H_u^0\right|^2 +
\left|H_u^+\right|^2 - \left|H_d^0\right|^2 -
\left|H_d^-\right|^2\right)^2
+\frac{g_2^2}{2}\left|H_u^+ H_d^{0*} + H_u^0 H_d^{-*}\right|^2\nn \\
&&+m_{S}^2 |S|^2
+\big( \lambda A_\lambda \left(H_u^+ H_d^- - H_u^0 H_d^0\right) S + 
\frac{1}{3} \kappa A_\kappa\, S^3 + \mathrm{h.c.} \big) \: . 
\label{eq:Vnmssm}
\eea
%
The set of (tree-level)
tadpoles\footnote{Note in advance that {\tt HOM4PS2} might fail to find all
possible minima of a given potential in the presence of a collection of
degenerate ones. To circumvent this problem in our numerical studies, we add
very small $SU(2)$-breaking terms to the tadpoles.}
now includes the one for the singlet (neutral) scalar field `$S$', over and
above those for the doublet Higgs fields. All the tadpoles now involve $\vs$,
the \vev~for the field `$S$'. Similar to the MSSM case presented in section
\ref{subsec:analysis-mssm}, with
$T_{i}=\frac{\partial V_{\mathrm{Higgs}}}{\partial v_i}$, where `$i$' stands
for the field with respect to which a partial derivative is taken, these
tadpoles are given by 
%
%
\begin{subequations}
\label{eq:tadpoles-nmssm}
\bea
T_{\higgsun} =  0 &=& g^2
v_u(v_u^2+v_{u^+}^2-v_d^2-v_{d^-}^2) + 2 (m_{H_u}^2+\lambda ^2 v_s^2)v_u 
+ g_2^2 v_{d^-}(v_u v_{d^-} + v_{u^+} v_d) \nn \\
 & & + \, 2\lambda^2 v_d (v_u v_d-v_{u^+}v_{d^-}) -2 \lambda v_d v_s(\alambda +
\kappa v_s) \: , 
\label{eq:Tu-nmssm}
\\ [8pt] 
T_{\higgsdn} = 0 &=& -g^2 v_d(v_u^2+v_{u^+}^2-v_d^2-v_{d^-}^2) + 2 (m_{H_d}^2
+\lambda ^2 v_s^2)v_d + g_2^2 v_{u^+}(v_u v_{d^-}+v_{u^+} v_d ) \nn \\ 
&&+ \, 2 \lambda^2 v_u (v_u v_d-v_{u^+}v_{d^-}) 
-2 \lambda v_u v_s(\alambda + \kappa v_s) \: ,
\label{eq:Td-nmssm}
\\ [8pt] 
T_{\higgsup} = 0 &=& g^2 v_{u^+}(v_u^2+v_{u^+}^2-v_d^2-v_{d^-}^2) 
+ 2 (m_{H_u}^2+\lambda ^2 v_s^2)v_{u^+} +
g_2^2 v_d(v_u v_{d^-}+v_{u^+} v_d ) \nn \\ 
&& - \, 2\lambda^2 v_{d^-} (v_u v_d-v_{u^+}v_{d^-}) 
-2 \lambda v_{d^-} v_s(\alambda + \kappa v_s) \: ,
\label{eq:Tup-nmssm}
\\ [8pt] 
T_{\higgsdm} =0 &=& -g^2 v_{d^-}(v_u^2+v_{u^+}^2-v_d^2-v_{d^-}^2) 
+ 2 (m_{H_d}^2+\lambda ^2 v_s^2)v_{d^-} + g_2^2 v_u(v_u v_{d^-}+v_{u^+} v_d ) \nn \\ 
&& - \, 2\lambda^2 v_{u^+} (v_u v_d-v_{u^+}v_{d^-}) 
+2 \lambda v_{u^+} v_s(\alambda + \kappa v_s) \: ,
\label{eq:Tdm-nmssm}
\\ [8pt] 
T_S = 0 &=& \lambda^2 v_s(v_u^2+v_{u^+}^2+v_d^2+v_{d^-}^2) 
+ 2 v_s(m_S^2 + \kappa \akappa v_s + 2
\kappa^2 v_s^2) \nn\\ 
&&-2\lambda (\alambda + \kappa v_s)(v_u v_d-v_{u^+}v_{d^-}) \: ,
\label{eq:Ts-nmssm}
\eea
\end{subequations}
where $\vu$, $\vd$, $v_{u^+}$ and $v_{d^-}$ are as defined in section
\ref{subsec:analysis-mssm} and $\vs={\mueff \over \lambda}$.
%
As before, we solve for the squared soft masses (tree-level) for the neutral 
Higgs states (including the singlet scalar) at the DSB vacuum. These are given
by
%
%
\begin{subequations}
\label{eq:nmssm-soft-mass2}
\bea
m_{H_d}^2 &=&- \mueff^2 - \lambda^2 \vu^2 - \frac{g_1^2+g_2^2}{4} (\vd^2-\vu^2)+
\mueff (A_\lambda+\kappa \vs) \tan\beta \: ,
\label{eq:tadpole-d}
\\
m_{H_u}^2 &=& -\mueff^2 - \lambda^2 \vd^2 - \frac{g_1^2+g_2^2}{4}
(\vu^2-\vd^2)+\mueff (A_\lambda+\kappa \vs)\cot\beta \; , 
\label{eq:tadpole-u} \\
m_S^2 &=& -\kappa A_{\kappa} \vs - 2\kappa^2 \vs^2 - \lambda^2 (\vd^2+\vu^2) +2
\lambda \kappa \vu \vd +\lambda \frac{\vu \vd}{\vs} A_{\lambda} \; .
\label{eq:tadpole-s}
\eea
\end{subequations}
%
By substituting the squared soft masses from equation \ref{eq:nmssm-soft-mass2}
into the potential of equation \ref{eq:Vnmssm}, one finds the expression for the
tree-level depth of the DSB vacuum as
%
\bea
V^{^{\mathrm{DSB}}}_\mathrm{Higgs}|_{_\mathrm{tree}} 
&=& -\kappa^2 \vs^4 -\frac{1}{3} \kappa A_{\kappa} \vs^3 
- \lambda^2 \vs^2 (\vd^2+\vu^2) -
\lambda \vs \vd \vu (A_\lambda+2 \lambda \vs) \nonumber \\
& & -\frac{g_1^2+g_2^2}{8} (\vd^2 -\vu^2)^2 -\lambda^2 \vd^2 \vu^2 \; .
\label{eq:depth-dsb-nmssm}
\eea
%
As earlier, to find possible deeper minima, we look for some flat directions in 
the field space. The $D$-flat directions are given by the same set of equations
as in equation \ref{eq:flat-dir}. In addition, there is now an $F$-flat
direction given by
%
\beq
\la (\vuplus \vdminus - \vu \vd) + \ka \vs^2 = 0 \; .
\label{eq:fflat-nmssm}
\eeq
%
A first study exploiting this flat direction has recently been discussed in reference \cite{Krauss:2017nlh}.
Along this $F$-flat direction the tadpole equation for the singlet scalar field 
`$S$' has the following two independent solutions
%
%
\begin{subequations}
\bea
\vs&=&0 \; ,
\label{subeq:solone} \\
\vu^2 +\vd^2+\vuplus^2 +\vdminus^2&=& {-2 \over \lambda^2} \left[ m_S^2 + \kappa
\vs \left( \akappa - \alambda + \kappa \vs \right) \right] \: .
\label{subeq:soltwo}
\eea
\end{subequations}
%
%
Solution \ref{subeq:solone}, when plugged in into the tadpole conditions
$T_{\higgsdn}=0$ and $T_{\higgsdm}=0$, gives $\vd=\vdminus=0$ as a trivial
possibility. However, such a solution does not yield a CB minimum since the only
non-vanishing charged Higgs \vev~$\vuplus$ can now be rotated away in the
presence of a non-vanishing $\vu$. Thus, by feeding $\vd=\vdminus=0$~ to the
tadpole conditions $T_{\higgsun}=0$ and $T_{\higgsup}=0$, we find, at the CC
minimum,
\beq
v_u^2+v_{u^+}^2 = -2 \frac{m_{H_u}^2}{g^2} \: ,
\label{eq:nmssm-circle-1}
\eeq
with its depth given by
\beq
V_{\mathrm{Higgs}(u)}^{\mathrm{CC}}= {-\mhiggsu^4 \over g^2} \: .
\label{eq:depth-cb-nmssm}
\eeq
%
Clearly, the value of the potential at this CC minimum is negative. Also, for
larger values of $\mhiggsusq$, the potential could turn out to be deeper than
the DSB vacuum at the tree-level itself, a possibility that is in clear contrast
to the MSSM case. This is intimately connected to the magnitude of $\mueff$ as
can be seen from equation \ref{eq:tadpole-u}. Note that we would have arrived at
the corresponding set of relations involving $\vd$, $\vdminus$ and $\mhiggsd$
had we, instead, chosen to plug in the first solution ($\vs=0$) into the tadpole
conditions $T_{\higgsun}=0$ and $T_{\higgsup}=0$, i.e.,
\beq
v_d^2+v_{d^-}^2 = -2 \frac{m_{H_d}^2}{g^2} \: ,
\label{eq:nmssm-circle-2}
\eeq
with its depth given by
\beq
V_{\mathrm{Higgs}(d)}^{\mathrm{CC}}= {-\mhiggsd^4 \over g^2} \: .
\label{eq:depth-cb-nmssm-2}
\eeq
%

A deeper CC minimum could also appear along a direction $v_s \neq 0$, with all
other \vevs~set to zero, since such a configuration is always a solution to the
tadpoles. The depth of such a minimum is given by
\beq
V_{\mathrm{Higgs}}^{\mathrm{CC}} \Big\rvert_{\vs \neq 0}
  = v_s^2 m_S^2+\kappa ^2 v_s^4+\frac{2}{3} \kappa \akappa  v_s^3  \; .
\label{eq:depth-cb-nmssm-3}
\eeq
This can give rise to two non-zero minima with
\bea
v_s=\Bigg\{\frac {-(A_ {\kappa} + \sqrt {A_ {\kappa}^2 - 8 m_S^2})} {4\kappa}, \; 
\frac {-(A_ {\kappa} - \sqrt {A_ {\kappa}^2 - 8 m_S^2})} {4\kappa} \Bigg\} \; .
\label{eq:singlet-sol}
\eea
The corresponding depths can be shown to possess non-negative potential values
if one of the following conditions are satisfied \cite{Beuria:2016cdk}
\beq
A_ {\kappa} < -3\sqrt {A_ {\kappa}^2 - 8 m_S^2} \quad \mathrm{or} \quad  A_ {\kappa} > 
3\sqrt {A_ {\kappa}^2 - 8 m_S^2} \;.
\label{eq:singlet-ineq}
\eeq
As we will see later, in the presence of such a CC minimum with non-negative
potential (and hence not so deep), a CB minimum could eventually turn out to be
the global minimum (the effective panic vacuum) of the potential.

We now turn to a possible CB minimum. Its presence is conveniently studied in
the rotated basis (introduced in equation \ref{eq:field-rotation}) with 
$\vuplus^\prime=v_d^\prime=0$ ~or~ $\vdminus^\prime=v_u^\prime=0$. Since the
$F$-flat direction mentioned in equation \ref{eq:fflat-nmssm} yields $v_s=0$ as
a solution, this is consistent with the direction
$\vuplus^\prime \vdminus^\prime =v_u^\prime v_d^\prime$. Choosing    
$\vdminus^\prime=v_u^\prime=0$, we obtain the following solutions for
$v_d^\prime$ and $\vuplus^\prime$ from the tadpole conditions in equation
\ref{eq:tadpoles-nmssm}:
%
%
\begin{subequations}
\label{eq:nmssm-cb-vevs}
\bea
|v_d^\prime| = \frac{\sqrt{-(g_1^2+g_2^2) \mhiggsdsq + (g_2^2-g_1^2) \mhiggsusq}}{g_1 g_2}
\label{subeq:vdprime} \; , \\ [10pt]
%
%
|\vuplus^\prime| = \frac{\sqrt{-(g_1^2+g_2^2) \mhiggsusq + (g_2^2-g_1^2)
\mhiggsdsq}}{g_1 g_2} \; .
\label{subeq:vuplusprime} 
\eea
\end{subequations}
%
%
\vskip 6pt
\noindent
The depth of the potential is given by
\vskip 6pt
\beq
V_{\mathrm{Higgs}}^{\mathrm{CB}}= 
 -\left[ \frac{g_1^2 (\mhiggsd^2+\mhiggsu^2)^2+g_2^2 (\mhiggsd^2-\mhiggsu^2)^2}{2 g_1^2
g_2^2} \right] \; ,
\label{eq:nmssm-cb-depth}
\eeq
which is clearly always negative. For the \vevs~in equation \ref{eq:nmssm-cb-vevs}
to be real, one requires the following sets of inequalities to hold
simultaneously:
%
\begin{subequations}
\label{eq:msq-constraints}
\bea
\frac{g_1^2+g_2^2}{ g_2^2 - g_1^2}\mhiggsusq &<& \mhiggsdsq  <
\frac{g_2^2-g_1^2}{ g_1^2 + g_2^2}\mhiggsusq \; ,
\label{eq:msq-constraints-first}
\\ [10pt]
\mhiggsusq &<& 0, \quad  \mhiggsdsq < 0 \; .
\label{eq:msq-constraints-second}
\eea
\end{subequations}
%
\vskip 5pt
\noindent
For typical values of $g_1$ and $g_2$, the inequality in equation
\ref{eq:msq-constraints-first} approximately reduces to 
\bea
2 \mhiggsusq< \mhiggsdsq<0.5\mhiggsusq \; .
\eea
At this point, using soft mass-squared terms mentioned in equations
\ref{eq:tadpole-d} and \ref{eq:tadpole-u}, for $\tanb > \sqrt{2}$ and
$\mueff>0$, we obtain the following approximate inequality:
\bea
 & v_0^2 \frac{\tanb}{\mueff(1+\tansqb)} \left(\la^2-0.4
\frac{\tansqb-1}{\tansqb-2}\right) - \left(\frac{\tanb}{\tansqb-2}+
\frac{\ka}{\la}\right)\mueff \;\;  < & \alambda  \nn  \\
& < \; v_0^2 \frac{\tanb}{\mueff(1+\tansqb)}\left(\la^2-0.4
\frac{\tansqb-1}{2\tansqb-1}\right)+\left(\frac{\tanb}{2\tansqb-1}-
\frac{\ka}{\la}\right)\mueff
\; , 
%
\label{eq:cb-inequality-1}    
\eea
\vskip 5pt
\noindent
and for  $1< \tanb < \sqrt{2}$ and $\mueff>0$, similarly, we find
\bea
\alambda<
v_0^2 \frac{\tanb}{\mueff(1+\tansqb)}\left(\la^2-0.4
\frac{\tansqb-1}{2\tansqb-1}\right)+\left(\frac{\tanb}{2\tansqb-1}- \frac{\ka}{\la}\right)\mueff 
\; .
\label{eq:cb-inequality-2}   
\eea
\vskip 5pt
\noindent
The expressions for the lower and/or the upper limits of the inequalities in
equations \ref{eq:cb-inequality-1} and \ref{eq:cb-inequality-2} swap their
positions for $\mueff < 0$. From these two equations, it is clear that appearance
of a CB minimum explicitly depends on the set of four parameters, i.e.,
\{$\alambda$, $\tanb$, $\ka \over \la$, $\mueff$\}. In addition, other
parameters such as $\akappa$ could work in tandem with a chosen set of these
four parameters to yield a consistent, non-tachyonic spectrum for the DSB vacuum
and in rendering the accompanying CB minimum global. It may also be noted that a
recent work \cite{Krauss:2017nlh} has addressed similar issues, guided by
$\tanb \approx 1$. We have checked that the inequality in
\ref{eq:cb-inequality-2}, in the limit $\tanb \to 1$, leads to observations that
agree with those of reference \cite{Krauss:2017nlh}. On the other hand, the
inequality in \ref{eq:cb-inequality-1} that refers to $\tanb > \sqrt{2}$,
explores further regions in the NMSSM parameter space where a deeper CB minimum
could pose a genuine threat to the stability of the DSB vacuum.

It is thus clear from the above discussion that both CC and CB minima that are
deeper than the DSB vacuum could appear simultaneously for a tree-level NMSSM
Higgs potential. We have further checked that the inequalities pertaining to the
CB minima (expressions \ref{eq:cb-inequality-1} and \ref{eq:cb-inequality-2})
imply those to be deeper than a CC minimum arising along the direction $\vs=0$
(always having a negative potential value; see equations \ref{eq:depth-cb-nmssm}
and \ref{eq:depth-cb-nmssm-2}). Note that the value of the potential at a CB minimum
is always negative (see equation \ref{eq:nmssm-cb-depth}).
In the presence of a CB minimum deeper than the DSB vacuum, 
the ``globality'' of the former is conservatively ensured if a CC minimum
along $\vs\neq0$ (equation \ref{eq:depth-cb-nmssm-3}, singlet-only direction) 
has a positive potential.	
The latter is achieved if $\akappa$
can be constrained as in equation \ref{eq:singlet-ineq}. The requirement of
non-tachyonic Higgs states further restricts the allowed ranges of $\alambda$
and $\akappa$.

In the present analysis, we deal with a somewhat broader region of the NMSSM
parameter space (when compared to reference \cite{Krauss:2017nlh}) that yields
deeper CB minima. This is facilitated by a relatively large radiative correction
to the potential. Thus, guided by equation \ref{eq:cb-inequality-1}, we expect
to find regions with a global CB minimum even for relatively low values of
$|\mueff|$ when ${\kappa \over \lambda} > 1$. In that case, $|\alambda|$ can be
larger than $|\mueff|$. Furthermore, we also take into account the
effect of thermal correction to the potential.
%
\subsection{Scanning of the NMSSM parameter space}
\label{subsec:scan-nmssm}
%
In this section, we present and discuss the results of our scan over the NMSSM
parameter space using the package \veva. This would shed light on regions of
the said parameter space with diverse kind of stability properties of the DSB
vacuum, without and with the inclusion of thermal contributions to the potential.
As has been noted in section \ref{subsec:scan-mssm}, we have tweaked \veva~so
as to find the most relevant `panic' vacuum. It may be mentioned here that we
have not seen any significant impact of such a modification in the MSSM case.
However, reference \cite{Krauss:2017nlh} has recently pointed out that the
issue becomes important in the NMSSM case, an observation on which we concur.

The analysis presented in section \ref{subsec:analysis-nmssm} prompts us to
divide the scan into two categories: (i) one which is suited for exploring a
deeper CB minimum guided by equations \ref{eq:cb-inequality-1} and
\ref{eq:cb-inequality-2} and for which we take fixed values of $\mueff$ and
$\kappa$ while $\alambda$ and $\akappa$ are varied and (ii) the other which is
tailored to find (mostly) a deeper CC minimum, guided by equation
\ref{eq:singlet-ineq}, for which $\akappa$ is kept fixed while $\alambda$,
$\mueff$ and $\kappa$ are varied. Note that, from equation \ref{eq:Vnmssm},
$\kappa$ and $\akappa$ govern the pure singlet contribution to the NMSSM Higgs
potential. Varying one or the other of these two parameters at a time would shed
light on how and to what extent the singlet sector carves out a CC or CB
minimum, possibly deeper than the DSB vacuum. For both cases, we hold $\lambda$
and $\tanb$ fixed at an optimal,  common set of values. We discuss these cases
in the next two subsections. All through, we keep track of the regions
compatible with the observed SM-like Higgs boson by using the packages
{\tt HiggsSignals} and {\tt HiggsBounds}. 
%
\subsubsection{Hunt for deeper charge-breaking minima: case with fixed
$\mueff$ and $\kappa$}
\label{subsubsec:scan-deeper-CB-nmssm}
In this subsection, we present the results of a random scan over a large region
in the $\alambda$--$\akappa$ plane keeping $\mueff$, $\kappa$ and some other 
parameters fixed at suitable values. A closer look at equation
\ref{eq:nmssm-cb-depth}, in conjunction with equations \ref{eq:tadpole-d} and
\ref{eq:tadpole-u}, would help us decide on the strategy for scanning the NMSSM
parameter space. As we have seen in section \ref{subsec:analysis-nmssm}, one of
the dangerous directions (an $F$-flat direction) along which a CB minimum could
appear is $\vs=0$. Its appearance, however, is facilitated by ensuring, to start
with, a shallower potential at the DSB vacuum. From the inequalities in
equations \ref{eq:cb-inequality-1} and \ref{eq:cb-inequality-2} that are
required to be satisfied for CB minima to occur, we find that somewhat large
values of $\la$ and $\mueff$ along with low values of $\tanb$ help. Hence we fix
$\la$ to a moderately large value of 0.7 and take $\tanb=2$. In addition, we
take a somewhat large value of $\kappa=1$ which, as discussed at the end of the
last subsection, help ensure ${\kappa \over \lambda} > 1$ thus enabling
exploration of a deeper CB minimum for somewhat larger values $|\alambda|$. To
demonstrate the latter, we choose two representative values of $\mueff$.
Furthermore, two sets of values of soft parameters ($\msQthree=\msUthree$ and
$A_t$) in the top squark sector (yielding small/large masses/mixings) are chosen
for the purpose. These amount to a varied extent of radiative correction to the
potential. Such choices are expected to alter the spans of the parameter plane
inflicted with panic vacua of both CC and CB types.
%
%
\begin{figure}[t]
\centering
\includegraphics[height=0.31\textheight, width=0.49\columnwidth ,
clip]{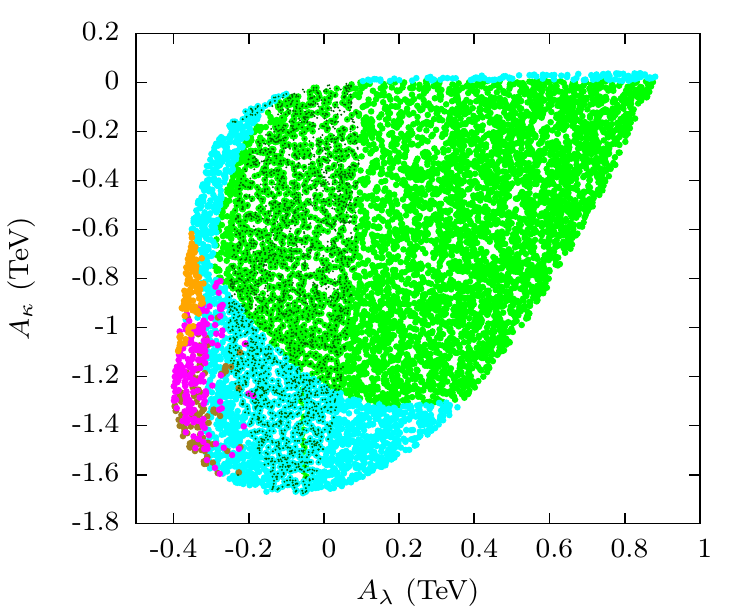}
\includegraphics[height=0.31\textheight, width=0.47\columnwidth ,
clip]{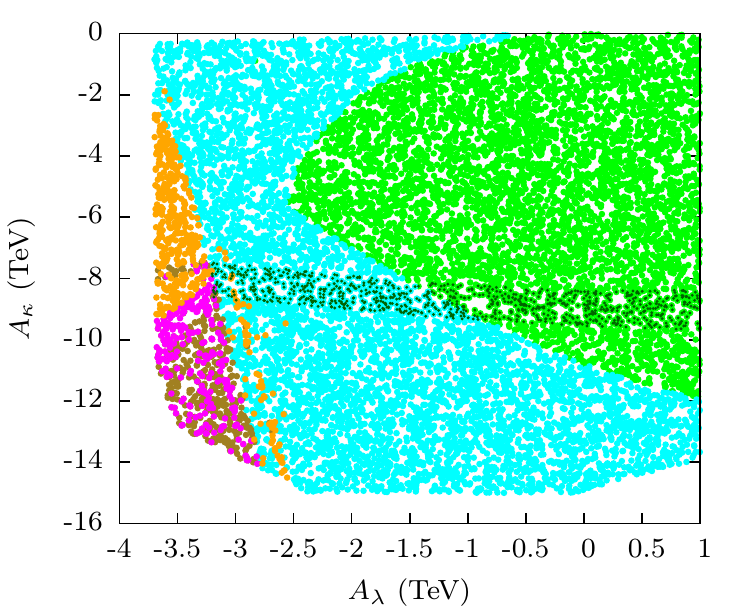}
\caption{Scatter plots obtained from a random scan over $\alambda$ and 
$\akappa$ and showing the stability pattern of the DSB vacuum in the
$\alambda$--$\akappa$ plane for $\mueff=300$ GeV (3 TeV) in the left (right)
plot. The color code is summarized in table \ref{tab:color-code}. Other
important fixed parameters are $\msQthree=\msUthree=750$ GeV, $A_t=0$,
$\lambda =0.7$ and $\kappa =1$.
}
\label{fig:low-stop-nmssm}
\end{figure}
%

In figure \ref{fig:low-stop-nmssm}, we present the region in the
$\alambda$--$\akappa$ parameter plane which possesses a DSB vacuum and may be
accompanied by a CC and/or a CB minimum which are/is deeper than the former. The
color code described in table \ref{tab:color-code} indicates only the presence
and nature (CC or CB) of such `panic' minima and not yet tells anything about
whether such panic minima are dangerous for the stability of the DSB vacuum.
The figure represents the case with low stop masses ($\approx 750$ GeV) and with
$\mueff=300 \, \mathrm{GeV} \, (3 \, \mathrm{TeV})$ for the left (right) plot.
It may be noted that the ranges of $\alambda$ and $\akappa$ are much larger for
the plot on the right with $\mueff=3$ TeV when compared to the left plot with
$\mueff=300$ GeV. This is since by increasing $\mueff$ and hence $\kappa v_S$,
one could accommodate large negative values of $\akappa$ and $\alambda$
consistent with a non-tachyonic Higgs spectrum. We now find a deeper CB minimum
appearing for such large negative values of $\alambda$ and $\akappa$. From
equation \ref{eq:cb-inequality-1} and \ref{eq:cb-inequality-2} we find that
$\alambda$ is governed by $-{\kappa \over \la} \mueff$ for small values of
$\tanb$. For this figure, ${\kappa \over \la }>1$ and hence the region of CB
minima appears around $\alambda < -\mueff$. Accordingly, this fixes the range of
$\akappa$ so that tachyonic states are avoided, as pointed out above. Note that
a flip of sign on $\mueff$ results in altered signs on both $\alambda$ and
$\akappa$ to find such regions with a deeper/global CB minimum.
%
%
\begin{table}[t]
\fontsize{9}{7}\selectfont
\centering
\begin{tabular}{|c||c|c|c|c|c|c|}
\hline
  & & & & & &  \\
Color & Green & Cyan & Orange & Brown & Magenta &
Dark-green \\
 & & & & & &  \\
\hline
 & & & & & & \\
Deeper vacua  & DSB only & Deeper CC  & Deeper CC & Deeper CB  & Deeper CB &
\textemdash \\
present &          & No deeper CB & Deeper CB & No Deeper CC & Deeper CC & \\
 & & & & & &  \\
\hline
 & & & & & & \\
Observation & DSB Global & CC Global & CC Global & CB Global & CB Global &
Allowed by \\
 & & & & & & Higgs data \\
 & & & & & &  \\
\hline
\end{tabular}
\caption{Color code used in figures \ref{fig:low-stop-nmssm} and
\ref{fig:large-stop-nmssm} to indicate the presence of minima deeper than the
DSB vacuum, their nature (CC or CB) and the one that is the global minimum of
the potential.}
\label{tab:color-code}
\end{table}
%
%

\begin{figure}[t]
	\centering
	\includegraphics[height=0.31\textheight, width=0.49\columnwidth ,
	clip]{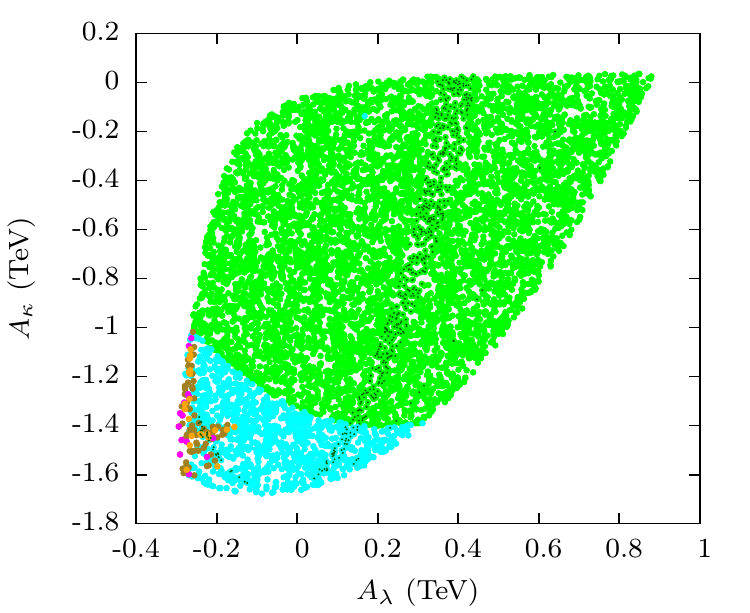}
	\includegraphics[height=0.31\textheight, width=0.47\columnwidth ,
	clip]{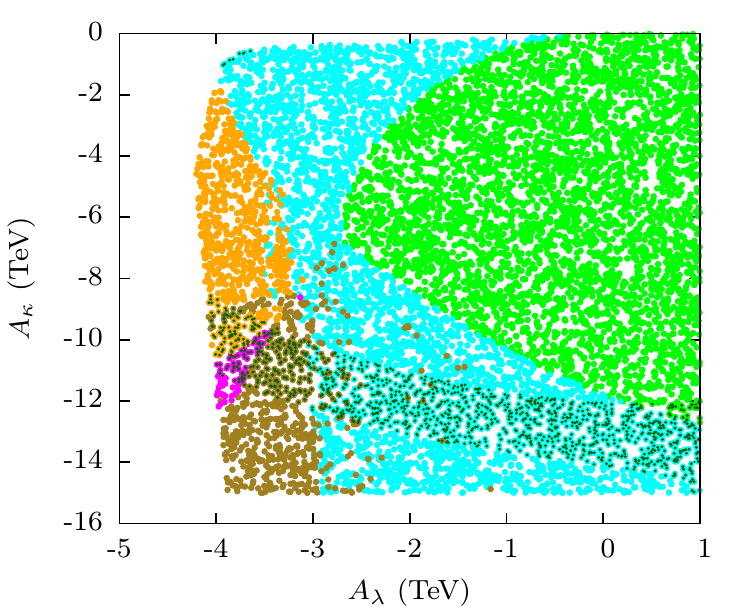}
	\caption{Same as in figure \ref{fig:low-stop-nmssm} but for
		$\msQthree=\msUthree=3$ TeV, $A_t=1$ TeV.}
	\label{fig:large-stop-nmssm}
\end{figure}

It would be now interesting to study the impact of a large radiative correction
to the potential. Such large corrections are easily achieved with larger values
of masses and mixings in the top squark sector. We thus fix
$\msQthree=\msUthree=3$ TeV and $A_t=1$ TeV, keeping other fixed parameters the
same as in figure \ref{fig:low-stop-nmssm}. Figure \ref{fig:large-stop-nmssm}
illustrates the case. The left plot (with $\mueff=300$ GeV) of figure
\ref{fig:large-stop-nmssm} hints a shrinking of the region featuring a deeper
CB vacuum (in orange, red and magenta) when compared to the corresponding one of
figure \ref{fig:low-stop-nmssm}. However, radiative effects are amplified for
larger values of $\mueff$ ($\sim \msQthree, \; \msUthree$), as can be seen by
comparing the right plots of these two figures.
\begin{figure}[t]
	\centering
	\includegraphics[height=0.31\textheight, width=0.48\columnwidth ,
	clip]{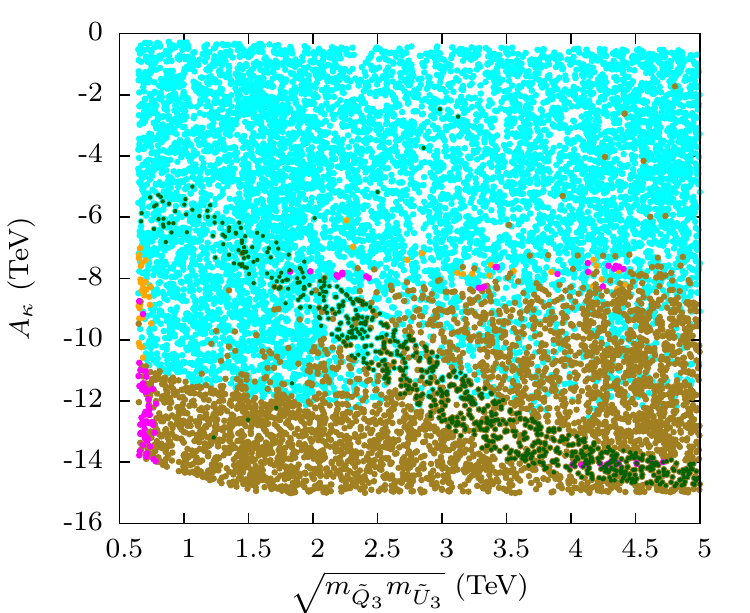}
	\caption{Same as in figure \ref{fig:low-stop-nmssm} but in the
		$\sqrt{\msQthree\msUthree} \,$--$\,\akappa$ plane. The fixed parameters are also
		as in figures \ref{fig:low-stop-nmssm} except for $\mueff=-\alambda=3$ TeV and
		$A_t=\msQthree=\msUthree$.}
	\label{fig:varying-stop-nmssm}
\end{figure}
%
\begin{figure}[t]
	\centering
	\includegraphics[height=0.31\textheight, width=0.49\columnwidth ,
	clip]{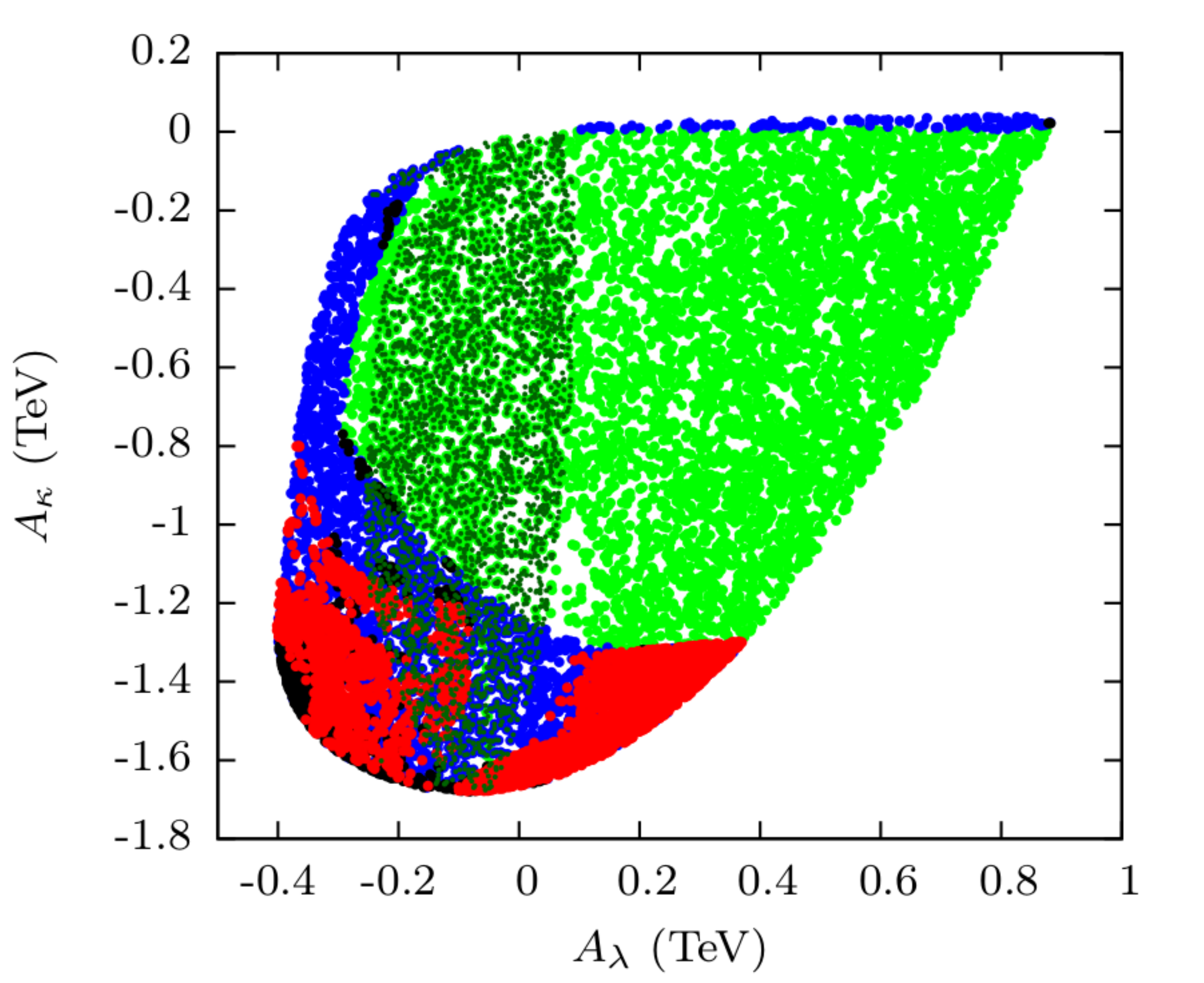}
	\includegraphics[height=0.31\textheight, width=0.47\columnwidth ,
	clip]{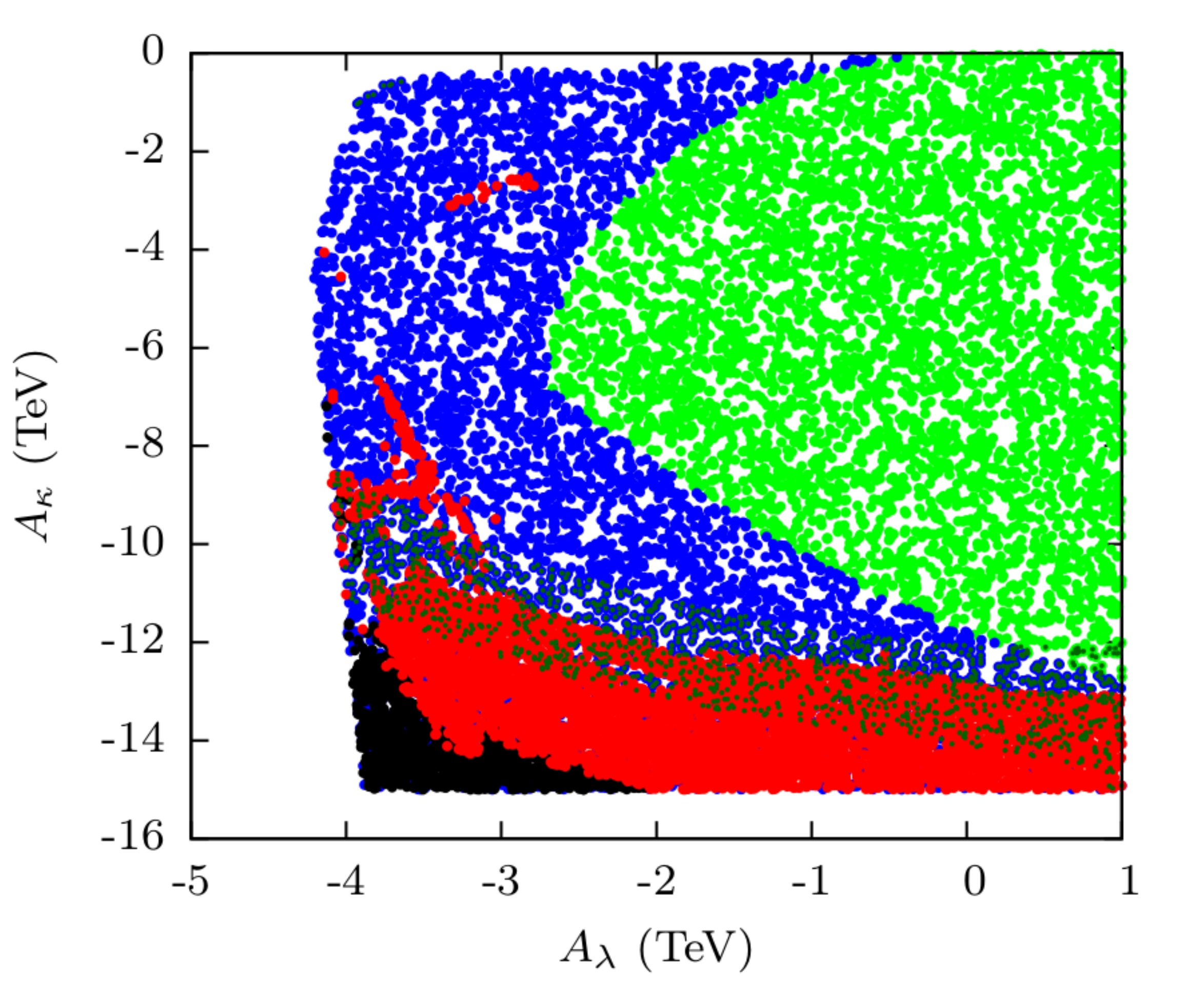}
	\caption{
		Scatter plots showing the stability status of the DSB vacuum in the
		$\alambda$--$\akappa$ plane in the presence of a deeper CB minimum. Color code
		in use are as adopted for figure  \ref{fig:mssm-mu-At-ccb}. The left (right)
		plot corresponds to the same set of data points as used in the left plot of
		figure \ref{fig:low-stop-nmssm} (right plot of figure
		\ref{fig:large-stop-nmssm}).}
	\label{fig:tunnel-nmssm-CB}
\end{figure}
It may be summarized from figures \ref{fig:low-stop-nmssm} and
\ref{fig:large-stop-nmssm} that a CB vacuum could turn out to be the global
minimum of the potential over an appreciable region of parameter space for
relatively large values of $\alambda$, $\akappa$, $\mueff$ and parameters in the
top squark sector and for $\tan\beta$ on the smaller side. Under the
circumstances, the global CB minimum could either be the lone deeper minimum
(in ``red'') or can be accompanied by a CC minimum which is shallower than it,
but still deeper than the DSB vacuum (in ``magenta''). Note that with
increased values of soft parameters in the top squark sector a global CB
minimum becomes increasingly compatible to observed Higgs boson properties.
Thus, a priori, such CB minima should be considered as dangerous for the
stability of the DSB vacuum. This warrants dedicated studies of vacuum
configurations of the potential by including \vevs~for the charged Higgs states.

The extent of the impact of radiative correction to the potential is further
investigated in the plane $\sqrt{\msQthree \msUthree}$ -- $\akappa$ as
illustrated in figure \ref{fig:varying-stop-nmssm}. Here, we consider $\mueff=3$
TeV and $\alambda=-3$ TeV. Clearly, the larger the (soft) masses for the top
squarks, the larger is the region in the parameter plane that possesses a global
CB minimum (without an accompanying deeper CC minimum (in red)) which remains to
be compatible with the observed Higgs data. Absence of a ``green'' region in
this plot only indicates that the DSB vacuum never becomes the global minimum of
the potential for such a set of NMSSM parameters and hence lives dangerously.

Finally, the fate of the DSB vacua in the presence of a deeper minimum (CC or
CB) is determined by calculating how fast could the former tunnel to the latter.
A viable DSB vacuum is either the global minimum of the potential or its
lifetime is comparable (or larger) than the age of the Universe. In figure
\ref{fig:tunnel-nmssm-CB} we profile such regions in the $\alambda$-$\akappa$
plane on the basis of stability (viability) of the DSB vacuum against tunneling
to a deeper minimum. For a straightforward comparison, we choose the left
(right) plot of this figure to correspond to the left plot of figure
\ref{fig:low-stop-nmssm} (right plot of figure \ref{fig:large-stop-nmssm}).
We observe that significant portions of the parameter plane characterized
primarily by large $\akappa$ could get ruled out due to fast tunneling of the
DSB vacuum which is triggered by thermal effects. Such a finding is in agreement
with the observations made in reference \cite{Beuria:2016cdk} but now is
generalized to the case where CB minimum deeper than the DSB vacuum is a
possibility. 

It may be noted here that unlike in the case of the MSSM, a deeper
CB minimum could arise without a conventional CCB minimum being triggered. 
CCB directions, that otherwise could be dangerous, can be avoided 
in the presence of singlet $vevs$ since the latter might contribute positively
to the potential \cite{Ellwanger:1999bv, Kanehata:2011ei}.
In fact, our {\tt Vevacious} scan mostly indicates the
region of parameter space for figure \ref{fig:varying-stop-nmssm} 
to be CCB safe. We verify this by allowing for $vevs$ for
the stops in our {\tt Vevacious} analysis. 
For a rough understanding of the phenomenon, we impose the 
relevant (tree level) criteria mentioned 
in \cite{Ellwanger:1999bv, Kanehata:2011ei} on our {\tt Vevacious} outputs  
and find the above mentioned region is mostly CCB safe.
%
\subsubsection{Hunt for deeper charge-conserving minima: case with
a fixed $\akappa$}
\label{subsubsec:scan-deeper-CC-nmssm}
In this subsection, we present the results of a random scan in the 
$\alambda$--$\mueff$ plane keeping $\akappa$ fixed but allowing $\mueff$ and
$\kappa$ to vary over moderate ranges, as would suffice for the purpose.
We take $\akappa=-1.5$ TeV and set $\msQthree=\msUthree=1$ TeV with $A_t=0$.
We stick to the choice of $\lambda=0.7$ and $\tanb=2$ made in the previous
section. The scan is already subjected to the scrutiny of {\tt HiggsSignals}
and {\tt HiggsBounds}. Hence the results presented would be straightaway
compatible with the observed SM-like Higgs boson.

\begin{figure}[t]
\centering
\includegraphics[height=0.31\textheight, width=0.49\columnwidth ,
clip]{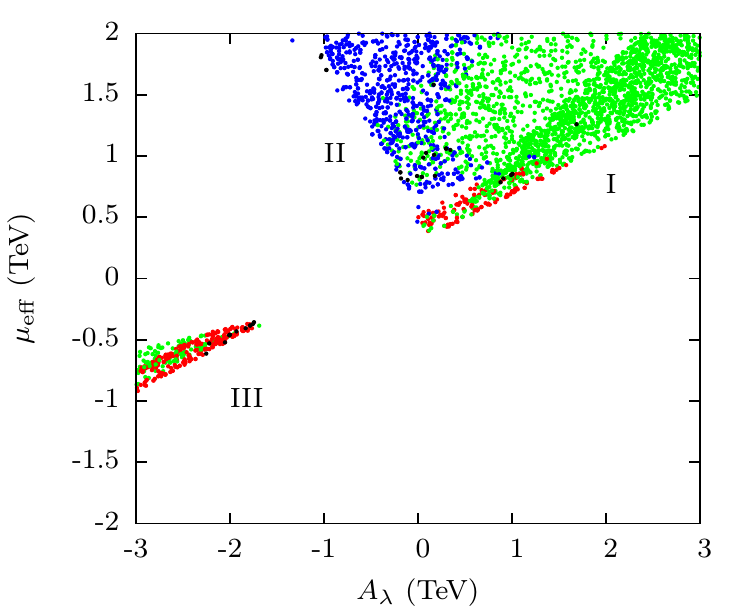}
\includegraphics[height=0.31\textheight, width=0.49\columnwidth ,
clip]{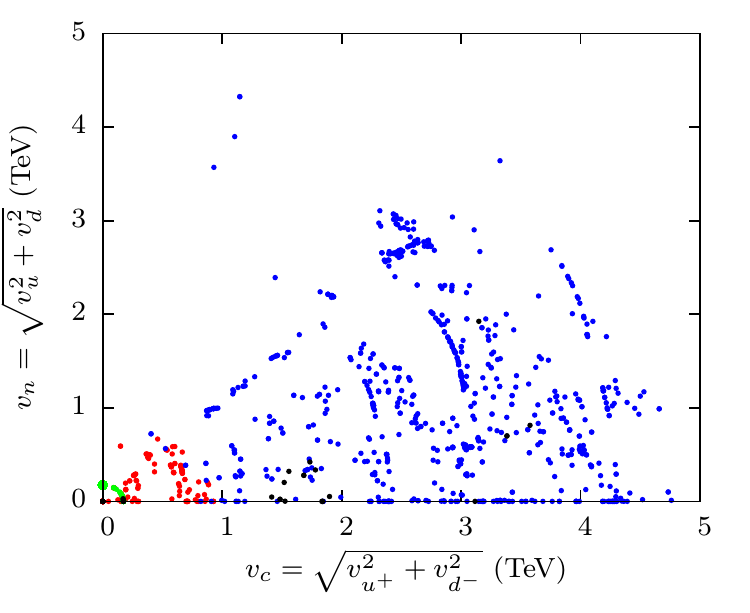}
\caption{
Scatter plots showing stability status of the DSB vacuum in the
$\alambda$--$\mueff$ plane (left) and in the $v_c$--$v_n$ plane (right).
Color code employed is as for figure \ref{fig:tunnel-nmssm-CB}. Ranges of 
various NMSSM parameters that are randomly scanned over are
$|\mueff| < 2$ TeV, $|\kappa|<0.75$ and $|\alambda|< 3$ TeV while the fixed
NMSSM parameters are $\lambda=0.7$, $\tan\beta=2$, $\msQthree=\msUthree=1$ TeV,
$A_t=0$ and $\akappa=-1.5$ TeV. All data points pass the constraints coming from
{\tt HiggsSignals} and {\tt HiggsBounds}.
}
\label{fig:nmssm-Alam-mueff-hb-hs}
\end{figure}

In the left plot of figure \ref{fig:nmssm-Alam-mueff-hb-hs} we illustrate the
stability status of the DSB vacuum in the $\alambda$--$\mueff$ plane. The ranges
of the parameters that are made to vary are indicated in the figure caption.
%
%
%
%
Regions I and II correspond to $\kappa > 0$ whereas region III has $\kappa < 0$.
Region I is found to have a stable DSB vacuum (in green) for larger values of
$\alambda$.
This can be understood by looking at the $\alambda$-dependent term in equation
\ref{eq:Vnmssm}. Furthermore, the same term predicts that the situation could 
change dramatically if $\alambda$ and $\mueff$ carry a relative sign since the
DSB vacuum could then become shallower relative to other non-DSB minima of the
potential that might be present. Region II represents such a situation. However,
the DSB vacuum is found to be mostly long-lived (in blue) over this region. We
observe that metastable DSB vacua could also appear for values of
$|\mueff| <1$ TeV which are eventually found to be thermally unstable (in red)
thanks to a moderately large value of $\akappa$ \cite{Beuria:2016cdk} that we
use. It may be noted that for such regions, $\mueff$ and $\alambda$ carry the
same sign. From what we learn from the previous section, these are unlikely to
be the CB minima and are merely the deeper (and dangerous) CC minima. On the
other hand, the metastable (blue) points in region II could have either kind of
minima. Region III has got negative $\kappa$ and hence requires both $\alambda$
and $\mueff$ to be negative as well to ensure a non-tachyonic Higgs spectrum.

A corroborative insight into nature of these dangerous vacua can be drawn from
the right plot of figure \ref{fig:nmssm-Alam-mueff-hb-hs}. This is a scatter
plot projecting the data points of the left plot in the $v_c$--$v_n$ plane,
where $v_c=\sqrt{\vuplus^2 + \vdminus^2}$ and $v_n=\sqrt{\vu^2+\vd^2}$. As has
been discussed in section \ref{subsec:analysis-mssm}, CC vacua could only appear
along the $D$-flat direction mentioned in equation \ref{eq:flat-dir-mssm-1}.
Deeper vacua that mostly conserve charge form circular patterns which can be
understood by looking at equations \ref{eq:nmssm-circle-1} and
\ref{eq:nmssm-circle-2}. This is further corroborated by a vanishing photon mass
arising with such a system of \vevs. A small arc of radius
($v_c^2+v_n^2 \approx 174$ GeV) in green, close to the origin, represents an
equivalent set of DSB vacua connected via $SU(2)$ transformations. Just beyond
this, a narrow belt in red represents the \vev~combinations leading to deeper
minima which make the DSB vacuum for each case thermally unstable. Further away
from the origin, a metastable DSB vacuum survives tunneling and becomes
long-lived (in blue).
%
\section{Conclusions}
\label{sec:conclusions}
%
\noindent
In this work, we have studied the possibilities and the implications of a
spontaneous breakdown of charge, triggered by the charged Higgs states acquiring
\vevs, in popular SUSY scenarios like the MSSM and the NMSSM.

It has been known for some time that in a generic 2HDM, in the presence of a
charge-conserving minimum, the tree-level Higgs potential cannot have a deeper 
minimum where charge breaks spontaneously. The MSSM being a SUSY extension of
such a scenario is not an exception. In fact, rigorous studies from the past
had already established that the tree-level MSSM potential cannot even have a
second minimum, either of CC or CB type, once it offers a DSB vacuum. In the
present work, we show that when quantum corrections are included in the MSSM
potential, a deeper CC minimum could arise along the $D$-flat directions
together with the DSB vacuum. A \veva-based thorough scan of the MSSM parameter
space reveals that such a deeper CC minimum is always accompanied by a
conventional CCB minimum. Furthermore, regions of the parameter space where such
a CC minimum appears are hardly ever compatible with the observed SM-like Higgs
boson. Hence, on both counts, such a deeper CC minimum cannot emerge as an
exclusive threat to the stability of the DSB vacuum. On the other hand, an
accompanying deeper CB minimum never shows up. These findings are further
corroborated by our alternate analysis using the {\tt FeynHiggs-Mathematica}
framework. The role of thermal correction to the potential is also discussed.

The situation is characteristically different in the NMSSM thanks to the
presence of a neutral, SM-singlet scalar field. Here, a deeper CB minimum along
with a CC one of a similar nature could already occur with the tree-level Higgs
potential. Thus, checking for the stability of the DSB vacuum becomes a rather
involved task, more so when radiative corrections to the potential are included.
The issue has also been recently studied in reference \cite{Krauss:2017nlh}.
We broadly agree with the inferences of that work. However, we further note that
there may be regions in the NMSSM parameter space, though a little remote to the
ones studied in reference \cite{Krauss:2017nlh}, where a deeper CB minimum could
arise. Unlike in the case of the MSSM, there may not be any accompanying deeper
CCB minimum. Hence such a CB minimum could pose a genuine threat to the
stability of the DSB vacuum and hence should not get overlooked.

We also demonstrate that, in the process, thermal corrections to the potential
are, in general, crucial and ignoring them could lead to grossly incorrect
information on the stability of the DSB vacuum.
%
\acknowledgments
%
JB is partially supported by funding available from the Department of Atomic 
Energy, Government of India for the Regional Centre for Accelerator-based 
Particle Physics (RECAPP), Harish-Chandra Research Institute. The authors like 
to thank Utpal Chattopadhyay for his participation during the initial stage of 
this work and for discussions. The authors acknowledge helpful discussions with
Jos\`e E. Camargo-Molina, Sven Heinemeyer, Wolfgang G. Hollik, Manuel E. Krauss,
Ben O'Leary and Florian Staub. They also acknowledge the use of the High
Performance Computing facility at HRI and thank Amit Khulve for technical
support.
%
%

%
\end{document}